\documentclass{article}
\usepackage[utf8]{inputenc}
\usepackage{geometry}
\usepackage{subcaption}
\usepackage{hyperref, url}
\usepackage{graphicx}
\usepackage{authblk}
\usepackage{natbib}
\usepackage{eurosym}
\title{Observed Patterns of Surface Solar Irradiance under Cloudy and Clear-sky Conditions}

\author[1]{\textbf{Wouter Mol} (wbmol@wur.nl)}
\author[1]{Bert Heusinkveld}
\author[1]{Mary Rose Mangan}
\author[1]{Oscar Hartogensis}
\author[1]{Menno Veerman}
\author[1]{Chiel van Heerwaarden}
\affil[1]{Meteorology and Air Quality Group, Wageningen University, Wageningen, The Netherlands}

\usepackage{graphicx}

\begin{document}

\maketitle
\footnotetext{Funding from the Dutch Research Council (NWO) (grant: VI.Vidi.192.068)}

\begin{abstract}
Surface solar irradiance varies on scales as small as seconds or meters due to scattering and absorption by the atmosphere.
Clouds are the main driver of this variability, but moisture structures in the atmospheric boundary layer and aerosols have an influence too, and depend on wavelength. 
The highly variable nature of solar irradiance is not resolved by most atmospheric models, yet it affects most notably the land-atmosphere coupling, which in turn can change the cloud field, and the quality of solar energy forecasting.
Spatially and spectrally resolved observational datasets of solar irradiance at such high resolution are rare, but they are required for characterising observed variability, understanding the mechanisms, and developing fast models capable of accurately resolving this variability. 
In 2021, we deployed a spatial network of low-cost radiometers at the FESSTVaL (Germany) and LIAISE (Spain) field campaigns, specifically to gather data on cloud-driven surface patterns of irradiance, including spectral effects, with the aim to address this gap in observations and understanding.
We find in case studies of cumulus, altocumulus, and cirrus clouds that these clouds generate large spatiotemporal variability in irradiance, but through different mechanisms and at difference spatial scales, ranging from 50 m to 30 km.
Spectral irradiance in the visible range varies at similar spatial scales, with significant blue enrichment in cloud shadows, most strongly for cumulus, and red enrichment in irradiance peaks, particularly in the case of semi-transparent clouds or near cumulus cloud edges.
Under clear-sky conditions, solar irradiance varies significantly in water vapour absorption bands at the minute scale, due to local and regional variability in atmospheric moisture.
In conclusion, observing detailed spatiotemporal irradiance patterns is possible using a relatively small, low-cost sensor network, and these network observations provide insights and validation for the development of models capable of resolving irradiance variability.\\


\end{abstract}

\section{Introduction}
Surface solar irradiance varies on scales as small as seconds or meters mainly due to clouds, influencing the coupled land-atmosphere system (i.e., the energy, water, and carbon cycle), atmospheric photochemistry, and solar energy production.
Cloud fields generate complex, high contrast spatial patterns that range from stationary to dynamically evolving depending on cloud type, cloud velocity, and cloud shape evolution. 
The spatiotemporal scales of surface irradiance variability are directly linked to cloud size distribution, which means a wide range of scales contribute to the total variance \citep{wood_distribution_2011, tabar_kolmogorov_2014, mol_reconciling_2023}.
Resulting heterogeneity in surface heat fluxes driven by these clouds can feed back to cloud development \citep{lohou_surface_2014, jakub_role_2017, veerman_case_2022}, and the highly variable nature of solar irradiance negatively impacts electricity grid stability and solar energy yield \citep{liang_emerging_2017, kreuwel_characterizing_2021, yang_review_2022}.
In addition to clouds, light scattering and absorption in the atmosphere by gas molecules and aerosols also result in changes in the light spectrum, which has implications for photosynthesis \citep{durand_diffuse_2021} and wavelength-dependant photovoltaic technologies \citep{dirnberger_impact_2015}.

Heterogeneity in surface irradiance is amplified by the three-dimensional nature of scattering, and thus redistribution, of solar radiation in the atmosphere. 
Cloud shadows are caused by the (partial) blocking of direct irradiance and are darker due to part of the light horizontally scattering to an area around the cloud shadow rather than directly in it.
When this scattered irradiance locally combines with unobstructed direct irradiance in a cloud-free area, it exceeds clear-sky irradiance, and potentially even extra-terrestrial irradiance \citep{yordanov_extreme_2015, gueymard_cloud_2017, cordero_surface_2023}.
Any such local increase of irradiance above clear-sky values is often referred to as 'cloud enhancement' \citep{gueymard_cloud_2017}.

As Earth system modelling moves to higher resolution and complexity, accurately resolving small scale variability in irradiance becomes increasingly important. 
While the physics of three-dimensional radiative transfer is well-known, the path light take through an atmosphere filled with liquid water and ice, aerosols, and gas molecules, over a partially reflecting surface, is highly complex. 
To add to the complexity, radiative transfer processes depend on wavelength, in particular in the case of absorption (e.g. due to ozone or water vapour) and Rayleigh scattering.

Qualitatively, the spatial patterns of irradiance that cloud fields generate can be visible by eye, e.g. from an aerial view or on the side of mountains, but a quantitative characterisation and explanation of how exactly they arise remain a challenge.
Spatial observations of surface irradiance at the scale of cloud shadow and enhancement patterns (10$^1$ - 10$^4$ m) \citep{mol_reconciling_2023}, with an adequate temporal resolution of 1 s or better \citep{tomson_fast_2010, yordanov_100-millisecond_2013}, are rare.
Many high quality solar irradiance observations exist, for example the Baseline Surface Radiation Network \citep{driemel_baseline_2018}, but while these adhere to the highest quality standards, they are effectively single point measurements at typically 1 minute resolution.
Examples of more dense networks are the 99 pyranometers network deployed during the HOPE campaign \citep{madhavan_shortwave_2016}, or the 17 photodiode pyranometers used by \cite{weigl_modelling_2012, tabar_kolmogorov_2014}, both deployed on $\sim$ 1 km$^2$ scale areas.
Photodiode pyranometers are fast responding ($<<$ 1 s) sensors, and thus capture the fastest atmospheric driven fluctuations in irradiance, but they do not provide spectral information.
None of these networks provide enough temporal, spectral, and spatial resolution to characterise surface irradiance heterogeneity at the scale of clouds, in part due to the high cost and operational burden of performing such measurements.

As for using modelling as a research tool, reproducing observed solar irradiance variability requires a fully resolved and realistic simulation of clouds, information about atmospheric composition and aerosols, and accurate 3D radiative transfer calculations using techniques such as Monte Carlo ray tracing.
In practise, the most detailed operational weather models operate at a resolution of approximately 1 km, too coarse to resolve clouds at the necessary scales.
Apart from that, the physics of radiative transfer is simplified to a two-stream approach (up and down) \citep{hogan_flexible_2018}, which by design means it cannot resolve cloud enhancement and will thus underestimate the contrast between shaded and sunlight surfaces. 
There are ways to improve upon two-stream methods by subgrid parameterisation of clouds and 3D radiative effects, such as SPARTACUS \citep{schafer_representing_2016} or ecRAD \citep{ukkonen_fast_2023} for operational weather models.
More accurately resolving 3D radiation is done in academic setups by for example \cite{veerman_case_2022}, who have coupled a 3D Monte Carlo ray tracer to a cloud resolving model, in an offline way as presented by \cite{gristey_influence_2022}, or using an alternative to ray tracing altogether as demonstrated by \cite{pincus_computational_2009}.
It can also be used to study specific mechanisms in a controlled manner, such as the effect of surface albedo \citep{villefranque_functionalized_2023}.
However, even in the best studied case of boundary layer shallow cumulus, one can question the realism with which large-eddy simulation (LES) can resolve such clouds \citep{romps_life_2021}.

Improving parameterised or optimised 3D radiative transfer calculations for finer scales depend on our understanding of 3D radiative transfer in the atmosphere and validation against observed variability.
To gather more observations of variability in total and spectral solar irradiance at cloud-scale, we have developed low-cost radiometers to be deployed in spatial network setups \citep{heusinkveld_new_2023}.
The design philosophy is similar to that of the APOLLO (Autonomous cold POoL LOgger) network \citep{kirsch_sub-mesoscale_2022}: autonomously running low-cost instruments optimised to accurately capture fluctuations rather than high accuracy single-point data.
With calibration against expensive, high-quality reference stations, these instruments give useful information at a fraction of the cost while also being flexible in their setup, necessary to densely cover and maintain a large area. 

In this paper, we describe the deployment, calibration, and first findings of a network of 20 to 25 radiometers at two field campaigns in 2021: FESSTVaL and LIAISE. 
Both campaigns were organised to observe and understand local meteorology, with FESSTVaL focused on sub-mesoscale variability and cold pools from convective storms, and LIAISE with the aim to observe the land-atmosphere coupling from irrigated crop fields to the regional scale in complex terrain.
The detailed observations of the atmosphere done by other groups at these campaigns are essential for understanding what we see in our own measurements.
Section \ref{sec:methods} covers the design and limitations of the radiometers, our network measurement strategy at both campaigns including a brief description of the weather, and a technique we use to construct spatial patterns from spatiotemporal data.
What follows are three sections of results, each with their own calibration and validation discussion.
First, we discuss three spatial patterns of broadband irradiance (Section \ref{sec:broadband}), followed by a study of how the visible part of the spectrum changes in these cases (Section \ref{sec:spectral}), and finally how irradiance varies in clear-sky conditions linked to water vapour variability (Section \ref{sec:wva}).
Conclusions and impacts of our results for solar irradiance variability research are discussed in Section \ref{sec:conclusions}.

\section{Methodology and campaigns}\label{sec:methods}

\begin{figure}[ht]
    \centering
    \includegraphics[width=\textwidth]{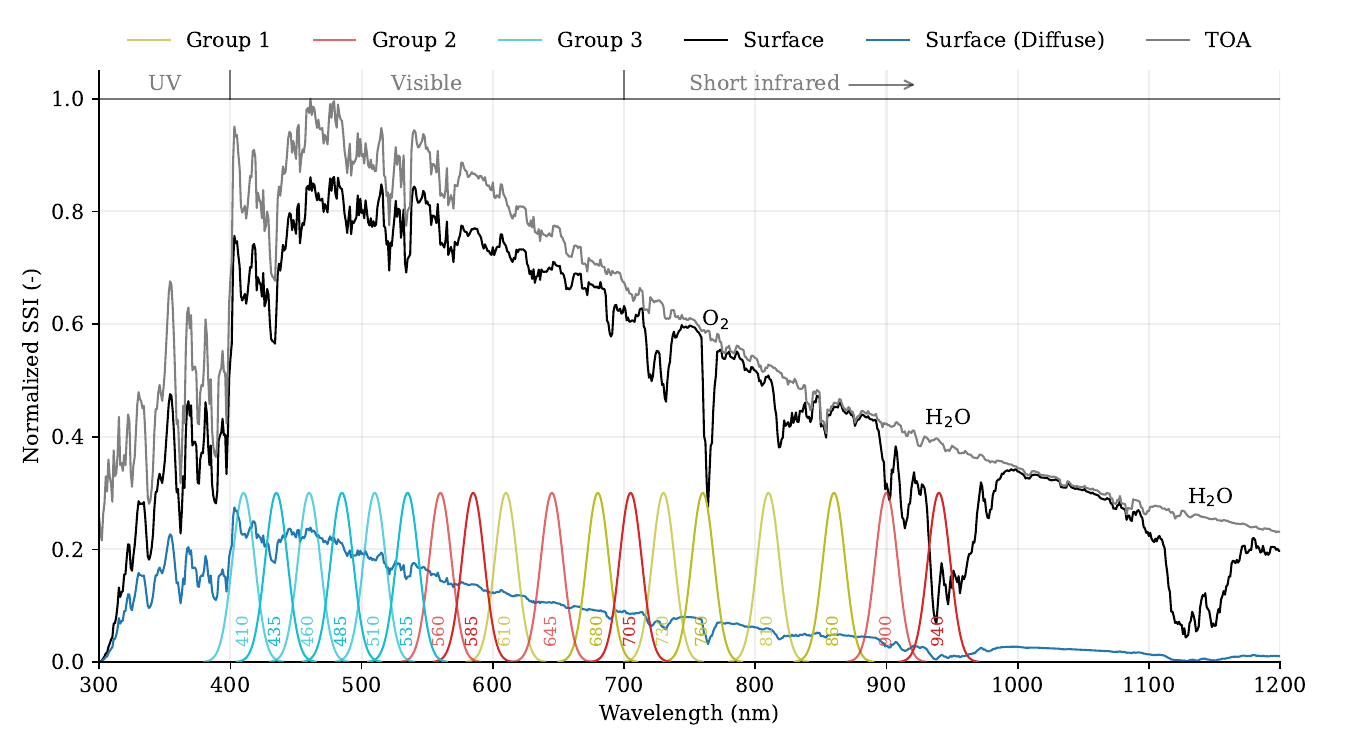}
    \caption{\textbf{Spectral solar irradiance and the radiometer wavelength bands.} Each radiometer band is color-coded according sensor subgroup. Normalized top of atmosphere and surface solar spectral irradiance are based on a clear-sky simulation using libRadtran \citep{emde_libradtran_2016} for 18 June 2021, 11:30 UTC, FESSTVaL campaign area in Germany (Section \ref{sec:modelinfo}). Spectra are smoothed with a 5 nm rolling mean for readability.}
    \label{fig:frostssi}
\end{figure}

\subsection{Solar irradiance sensor design}\label{sec:methodradiom}
The sensors we use have been specifically designed to capture the fastest cloud-driven fluctuations of incoming sunlight, and variations in the light spectrum induced by the atmospheric composition, clouds, aerosols, and vegetation.
At a sampling frequency of 10 Hz, the Fast Response Optical Spectroscopy Time synchronised instrument (FROST, \cite{heusinkveld_new_2023}), measures incoming shortwave irradiance at 18 wavelengths in the visible to near-infrared spectrum (410 to 940 nm). 
The locations of the 18 bands are detailed in Figure \ref{fig:frostssi}, which shows the response curves for the factory specification of 20 nm full-width half maximum, combined with a simulated solar spectrum of a mid-latitude summer day around solar noon.
The 18 bands are spread over 3 subgroups of 6 bands on the spectrometer chip, which are color-coded in the figure.
Since these three subsensors are spatially separated by about a centimetre (in a triad), we use a Teflon diffuser on top to equally distribute incoming sunlight.

Material costs for one sensor are under \officialeuro 200 in total, they are powered by their own small solar panel, and are all time-synchronised using a GPS chip. 
This makes them scalable and easy to deploy on tripods in field campaigns in flexible setups compared to conventional high quality (10 times or more expensive) pyranometers or spectrometers.
The low-cost design philosophy is a trade-off against accuracy compared to high-end instrumentation, but performance is good enough to capture and analyse spatial surface irradiance patterns driven by clouds and spectral signals of these variations.
We focus on the practical application of the sensor in this work, but a complete and technical reference with more use cases is described in \cite{heusinkveld_new_2023}.
Important sources of error are discussed next, which we either correct for or take into account in the analysis presented in Sections \ref{sec:broadband} to \ref{sec:wva}.

\subsection{Sources of measurement error}\label{sec:frosterror}
\subsubsection{Cosine response}
\begin{figure}[ht]
    \centering
    \includegraphics[width=\textwidth]{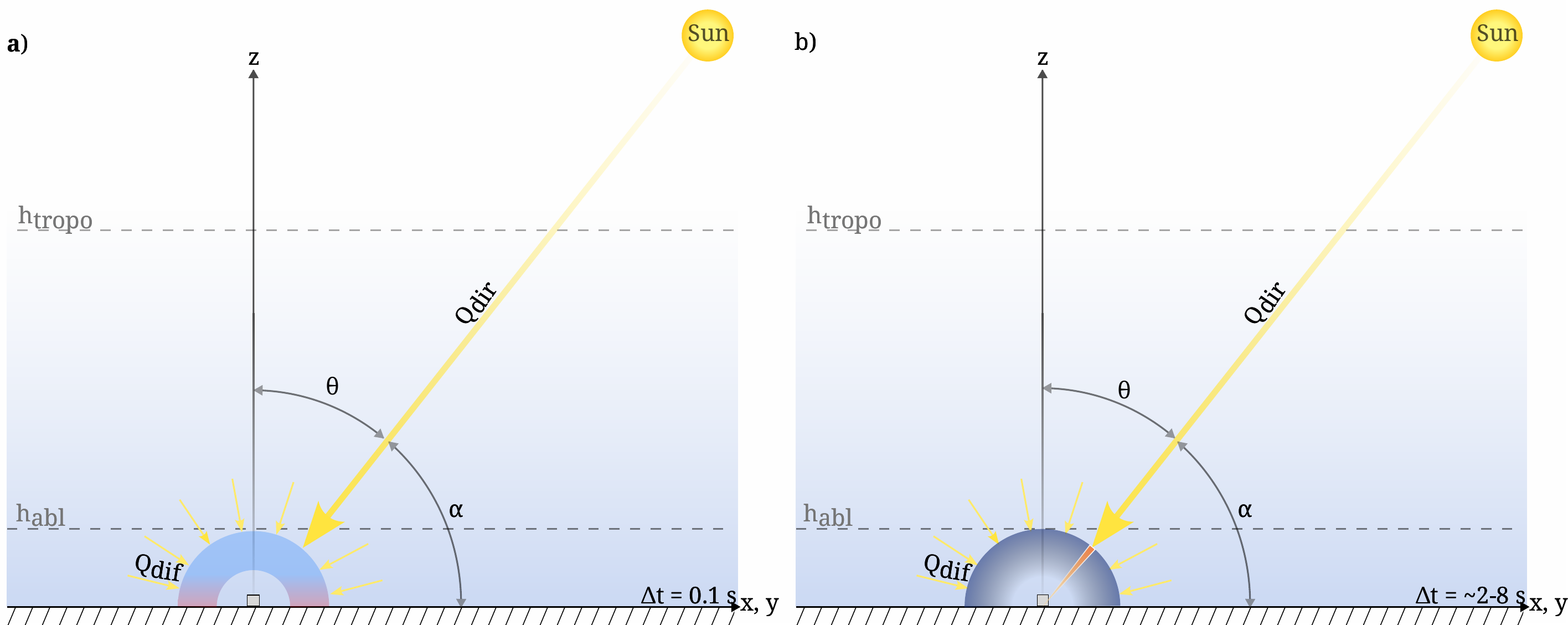}
    \caption{\textbf{Schematic of surface solar irradiance measurements} using FROST \textbf{(a)} and a suntracker with a pyranometer and pyrheliometer \textbf{(b)}, both located at gray square in the axis origin. The solar elevation angle $\alpha$ depicts the origin of direct irradiance Q$_{\mathrm{dir}}$, of which a fraction (typically 10 to 30\%, see Figure S1) scatters and results into diffuse irradiance Q$_{\mathrm{dif}}$, in this clear-sky example. FROST does not distinguish between Q$_{\mathrm{dir}}$ and Q$_{\mathrm{dif}}$, and misses part of the signal originating from low angles $\alpha$, illustrated by the red part of the semi circle in \textbf{(a)}. A suntracker measures Q$_{\mathrm{dir}}$ and Q$_{\mathrm{dif}}$ separately, illustrated by the 5$^{\mathrm{o}}$ degree arc and semi circle in \textbf{(b)}, and is less biased for small $\alpha$. The blue background gradient illustrates decreasing air density with height, h$_{\mathrm{tropo}}$ and h$_{\mathrm{abl}}$ are the approximate tropopause and atmospheric boundary layer heights. $\Delta$t denotes the resolved temporal resolution of each instrument.}
    \label{fig:scheme_rad}
\end{figure}

Given a solar zenith angle $\rm \theta$, the horizontally measured signal strength of a constant light source $\rm Q$ as function of $\theta$ would ideally be $\rm Q$ $\cdot$ cos($\theta$), see also Figure \ref{fig:scheme_rad}.
In practise, there is an increasing relative underestimation of irradiance for high $\theta$ in our instrument (Figure \ref{fig:scheme_rad}a), referred to as the cosine response, for which we correct in post processing.
While this cosine response correction is in principle a function of cos($\theta$) as well, the triad sensor design results in a unique response curve for each subsensor, despite the diffuser, which is also a function of the orientation of the sensor with respect to the sun. 
This means that the sensor orientation is important to keep constant throughout a measurement campaign, and ideally all sensors in a network are placed in the same orientation to minimise variations among sensors.
The variation in cosine response between subgroups is relevant for the usability of the ratio between spectral bands, of which we make use in Sections \ref{sec:spectral} and \ref{sec:wva}, as these bands can be on different subgroups.
This generally limits the use of spectral analyses to zenith angles of about 65 degrees or below.
Relative errors in spectrally integrated irradiance will also become pronounced at zenith angles of 75 or higher. 

\subsubsection{Build and placement consistency}
All instruments are hand-made, which leads to small imperfections or inconsistencies, such as the exact distance and position of the spectrometer from the diffuser.
It is also challenging to place and keep all sensors level within a tolerance of 0.5 degrees from day to day.
This results in measurable variations among instruments, more noticeable at high zenith angles, and imposes a limit on what is achievable through post processing.
While ideally there would be one universal calibration for all instruments, we find the best overall results when calibration is fine-tuned per individual instrument when possible.
This ultimately leads to a limitation in accuracy, which we quantify as the spread among sensors in a network.

\subsubsection{crosstalk}\label{sec:crosstalk}
Especially the third group of wavelength bands (Figure \ref{fig:frostssi}) experiences significant crosstalk, meaning a sensitivity to wavelengths $\lambda_{ct}$ outside the specified range of the band.
Under worst-case conditions (a flat spectrum of light plus a 10.6 mm Teflon diffuser) the 410, 435, and 460 nm bands derive $\sim$ 70 \% of their total signal from $\lambda$ $>$ 750 nm.
In reality, the solar spectrum is significantly less energetic for $\lambda_{ct}$ compared to 410 - 465 nm, effectively halving the crosstalk.
The bands of group 1 in Figure \ref{fig:frostssi} are all affected less than 10 \%.
The second group between 10 and 20 \%, except for 585 nm with $\sim$ 35 \% crosstalk. 
This issue is resolved in a new version of the instrument by using certain wavelength filters over the affected subsensors \citep{heusinkveld_new_2023}, but the sensor version used in this work is still without such filters.
Since the crosstalk effectively adds extra signal to be used when integrating spectra to broadband irradiance, we use all 18 bands in Section \ref{sec:broadband}.
For spectral analyses, we quantify the effect crosstalk has on measured changes in irradiance spectra and choose bands that are least affected. 
Response curves for each band and diffuser are shown, and available as supplementary material, in \cite{heusinkveld_new_2023}.

\subsubsection{Temperature sensitivity}
There are two components introducing a temperature sensitivity in the instrument, with no significant difference across wavelengths. 
A small, linear change in signal strength of -0.25\% per +10 K comes from the spectrometer itself. 
The teflon diffuser has a +2\% jump in transmittance from 20 to 21 $^{o}$C, and slowly declines at a rate of approximately -0.4\% per +10 K afterwards.
While 10 cm or 2 m air temperature is known for all measurements, it does not directly translate to Teflon diffuser and spectrometer temperature, making a temperature correction not trivial.
We use measured surface temperature $\mathrm{T_s}$ (10 cm) as a proxy for qualitative assessment of measurement quality, while keeping in mind $\mathrm{T_s}$ is still an underestimation.
The role of temperature in measurement accuracy is discussed further in Section \ref{sec:bbcalib} on broadband calibration.

\subsubsection{Factory calibration}
Because of limitations of the spectral quality in the sensor version used in this research, it is not easy to derive accurate measurements in W m$^{-2}$ nm$^{-1}$.
However, ratios between certain bands and changes therein contain valuable information and can be done in native sensor units, i.e., without calibration.
Sensor to sensor and wavelength band to band variations in factory calibration accuracy is generally $\pm$ 10 \%, with some outliers up to $\pm$ 20 \%, which also affects ratios between bands among sensors (Figure S2a).
Even though each sensor can be treated separately in some cases, it often helps to have homogeneous raw output among sensors for a given light signal.
We therefore homogenise the factory calibration as well as possible using a clear-sky periods where all sensors should measure the same, prior to performing any spectral analyses.
This reduces the spread to within 2 \% (Figure S2b), and produces a dataset labelled as 'precalibration', which is the starting point for spectral analyses presented in this study.

\subsubsection{Maintenance and quality control}
We performed irregular but frequent maintenance on the sensors during the field campaigns, usually early or late in the day, to check whether they were still running, level, and free of dirt or dust (birds or flies liked to sit on some particular sensors).
All data is provided with quality flags that mark data points with bad or unreliable data, which is mostly due to periods of sensor maintenance.
In addition, temporary displacement of sensors or the obscuring of direct sunlight by nearby objects at low solar angles (trees, crops, other instrumentation) is flagged.

\subsection{Measurement strategy at field campaigns}\label{sec:strategy}
We participated in two major field campaigns aimed at observing local to regional scale atmospheric dynamics and land-atmosphere coupling.
Two weeks in June 2021 during FESSTVaL (Field Experiment for Submesoscale Spatio-temporal Variability in Lindenberg, \cite{hohenegger_fesstval_2023}) in north-east Germany, and two weeks in July 2021 during LIAISE (Land surface Interactions with the Atmosphere over the Iberian Semi-arid Environment, \url{https://liaise.aeris-data.fr/}) in north-east Spain.
Differences between the two campaigns in climate, local atmospheric dynamics, time of year, and geographical location have offered a diverse range of solar irradiance conditions to observe through the network of sensors, resulting in a total of $\sim$ 4 weeks of spatial measurements.
The campaigns and sensor network measurement strategy are described next.

\subsubsection{FESSTVaL}
\begin{figure}[ht]
    \centering
    \includegraphics[width=0.9\textwidth]{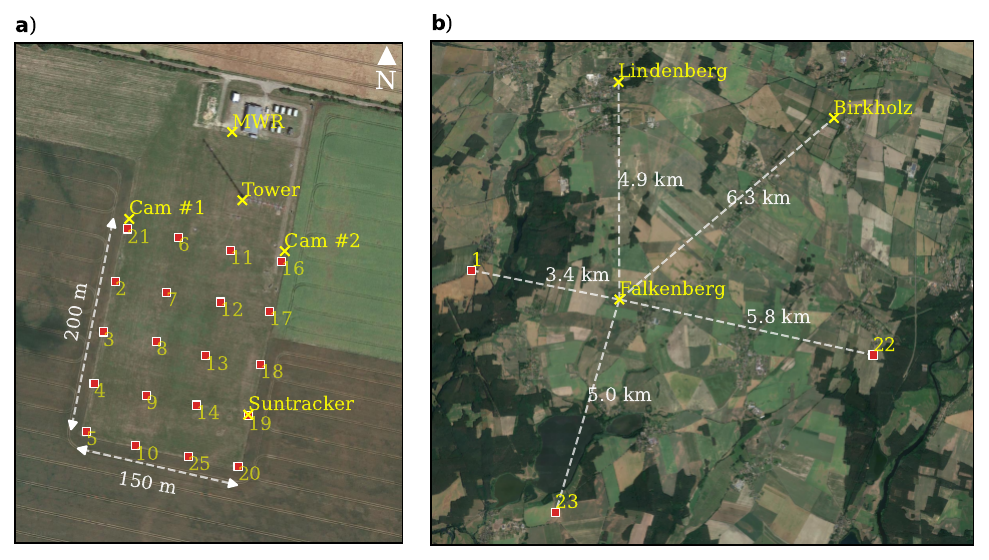}
    \caption{\textbf{Radiometer network measurement layout at FESSTVaL}. The 4 by 5 sensor grid at Falkenberg with a horizontal spacing of $\sim$50 meters is shown in \textbf{(a)}, together with the 98 meter tower, suntracker, two cloud cameras, and microwave radiometer (MWR) locations. Number labels are sensor IDs. Additional sensors set up in the FESSTVaL campaign area around Falkenberg are shown in \textbf{(b)}, and includes the location of the FESSTVaL supersites Lindenberg and Birkholz. Background satellite data: Google {\textcopyright} 2023.}
    \label{fig:gridf}
\end{figure}

During the FESSTVaL campaign, we deployed a network of 20 sensors in a simple, equidistant rectangular grid.
Our measurements took place between June 14 and June 29, 2021, at the Falkenberg supersite of the Deutscher Wetterdienst, $\sim$ 50 meters above sea level.
Figure \ref{fig:gridf}a shows the 4 by 5 sensor network layout with a 50 meter horizontal grid spacing. 
The choice of grid spacing is a combination of aiming for something that resembles the grid of a high resolution cloud resolving model, an a priori estimate of the required resolution to resolve shadow/sunlight transitions, practical constraints of the Falkenberg site, and the number of sensors we had available.
Technically, the grid spacing was 49 meters due to the constraints of rolling out the sensor network that is as little as possible obstructed by, or in the way of, other instrumentation on the field.

We deployed two consumer action cameras (with an on-board GPS clock) at the northern two grid corners to take time-lapse photos of the sky at a 5 second interval, so that we can relate the cloud field to surface irradiance.
These cameras were oriented up towards the sun in the south-east for the north-western camera, and south-west for the north-eastern camera.
Calibration of the sensors is done against the Falkenberg suntracker, equipped with high quality instrumentation and located in the south-east corner of the field.
Instrument locations relative to the grid are illustrated in Figure \ref{fig:gridf}a.

\begin{figure}[htb]
    \centering
    \includegraphics[width=\textwidth]{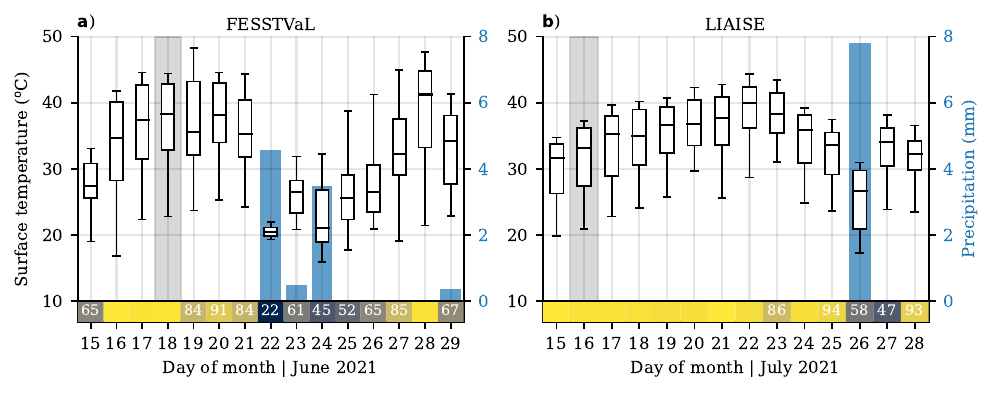}
    \caption{\textbf{Overview of weather variables most relevant for irradiance measurements at FESSTVaL and LIAISE}. All data is during day time (solar elevation angle $>$ 15$^{\mathrm{o}}$). Box plots are the surface temperatures at \textbf{(a)} Falkenberg and \textbf{(b)} Els Plans. The blue bars are accumulated precipitation during day time at the sensor network. The color shading at the bottom is the observed percentage of clear-sky irradiance (CAMS McClear), where darker colors indicate more cloudiness, labelled with numbers for values $<$ 95\%. Vertical gray bars at June 18 and July 16 are the reference cloud-free calibration dates used in Section \ref{sec:broadband}.}
    \label{fig:campaignwx}
\end{figure}

In addition to this setup, three sensors (1, 22, 23) were located several kilometers to the west, south, and east of the field (Figure \ref{fig:gridf}b), to capture part of the larger scale variability in the campaign area. 
Two other supersites are Lindenberg and Birkholz, and all supersites were equipped with microwave radiometers which measure integrated water vapour, used in this study, among other things.
There were many more instruments deployed, see \url{https://fesstval.de} and Hohenegger et al., (2023) for details.
Our two weeks at FESSTVaL featured one fully clear-sky day, one rainy day with thick cloud cover, but was otherwise characterised by many different (broken) cloud covers (Figure \ref{fig:campaignwx}a). 

\subsubsection{LIAISE}
\begin{figure}[ht]
    \centering
    \includegraphics[width=\textwidth]{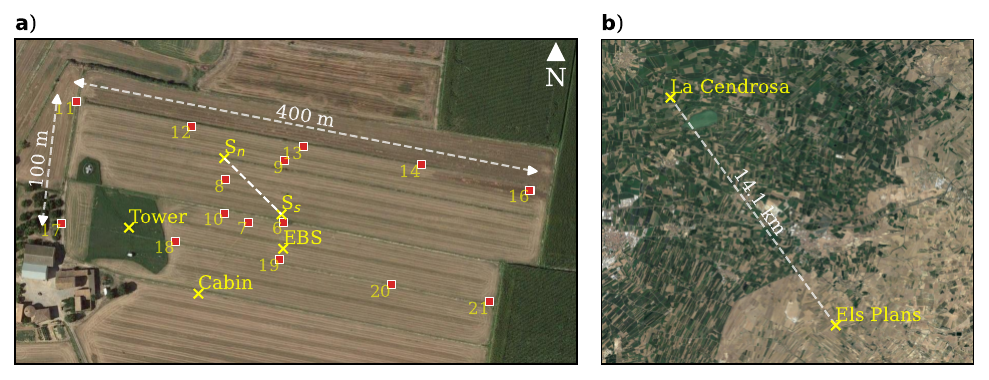}
    \caption{\textbf{Sensor network measurement layout at LIAISE}. The west-east oriented 5 by 2 sensor grid at La Cendrosa has a horizontal spacing of $\sim$100 meters, with an additional set of 5 sensors within the footprint of the scintillometer (S$_{\rm n}$ to S$_{\rm s}$), shown in \textbf{(a)}. The reference pyranometer is mounted on the EBS (energy balance station). La Cendrosa and Els Plans are shown in \textbf{(b)}, which illustrates the irrigated (green) versus non-irrigated (beige) area. Background satellite data: Google {\textcopyright} 2023.}
    \label{fig:gridl}
\end{figure}

Between 14 and 30 July 2021, we set up a network of sensors at the La Cendrosa site of the LIAISE campaign (\url{https://liaise.aeris-data.fr/}).
La Cendrosa is located within an irrigated part of an otherwise semi-arid region, with complex local, regional, and mesoscale dynamics \citep{mangan_surface-boundary_2023}.
The typical expected, and observed (see Figure \ref{fig:campaignwx}b), summertime weather in this region in north-eastern Spain is cloud-free, dry, and hot.
Specifically at La Cendrosa, due to local topography and sea breeze dynamics (locally called 'Marinada'), prevailing day time winds are westerly, shifting via  a southern sea breeze to easterly night time winds. 
While the goal is primarily to observe cloud-driven irradiance variability, frequent clear-sky days offer a good calibration opportunity and analyses in spectral variations due to day-to-day variations in aerosol and water vapour content.
We had a bit more space to set up the network compared to Falkenberg, so in an attempt to capture larger patterns, we decided on a grid spacing of 100 meters.
With the prevailing westerly daytime winds, we oriented the grid in a similar direction, in hopes of tracking cloud shadows and enhancements over a length of 400 meters, illustrated in Figure \ref{fig:gridl}a.
An additional group of radiometers was set up in footprint of the scintillometer, which in the context of this study gives additional resolution in the grid center, but makes the total network non-equidistant.

Two action cameras were mounted on the energy balance station (EBS), west and east oriented, but pointed straight ahead rather than at the sun to include a visual record of the vegetation growth and irrigation during the campaign period.
The EBS measures incoming broadband irradiance as part of the radiation balance measurements, which is used in this study as a calibration reference. 
Hourly boundary layer soundings were deployed at La Cendrosa during Intensive Observation Periods (IOPs), which we combine with hourly full troposphere soundings at the non-irrigated site Els Plans (Figure \ref{fig:gridl}b), located 14.1 km to the south east.
There are many more observations and sites within the campaign area, which can be found on \url{https://liaise.aeris-data.fr/}.

\subsection{Visualising spatial patterns}\label{sec:spatialinterp}
We find, as will be shown shortly, that cloud shadow and enhancement patterns often both exceed the network size and have details finer than the network resolution.
It is furthermore challenging to visualise a high amount of spatiotemporal data in a concise way without applying statistics.
For some figures, we therefore apply a data processing technique that makes use of the high temporal resolution and an estimate of the cloud velocity in order to increase the effective network size and spatial resolution.
Essentially, the following technique transforms temporal data to spatial data.
The technique assumes clouds retain their shape (analogous to Taylor's hypothesis of frozen turbulence) when moving over the measurement network (a schematic example is available in Figure S3).
For example, we can 'advect' the Falkenberg network (Figure \ref{fig:gridf}a) in space with a time step of 5 s and 5 m s$^{-1}$ cloud velocity for 10 steps, resulting in effectively a spatial network of 200 points (20 sensors $\times$ 10 steps) that spans $\sim$ 450 m (5 s $\times$ 5 m s$^{-1}$ $\times$ 10 steps + original network length of 200 m) in the advection direction and has approximately double the resolution (one step is 25 m).
The virtual 200 points can then be interpolated to a new equidistant grid for easier comparison and visualisation.
The main challenge with this technique is the determination of the cloud velocity vector, which we initially base on wind speed at cloud level from nearby soundings and ceilometer data, and then manually fine-tune to produce an as smooth as possible result.
Small changes in direction or speed quickly result in noisy results with artefacts from incorrectly placed data points reporting contradicting values.
Keeping the total integration time short minimises our violation of the static cloud shapes assumption, though some artefacts can remain.
The cloud velocity step fine-tuning needs to be repeated frequently, every 10 to 30 minutes or so, due to wind and cloud velocity (subtly) changing or simply varying among clouds.
Results are consistent when this exercise is repeated with selective omission of sensors.
We apply this technique to three distinct cases, first shown in Section \ref{sec:broadband}.

\subsection{Simulated solar position, clear-sky irradiance, and irradiance spectra}\label{sec:modelinfo}
In addition to observations, we require extra information about solar irradiance, mostly for the interpretation of measurements.
An estimate of clear-sky irradiance is required for identifying cloud enhancement events in the measurements, which we base on the globally available CAMS McClear product \citep{gschwind_improving_2019}.
This product is based on a radiative transfer model that calculates, for a cloud-free atmosphere, global horizontal (and diffuse) irradiance, given a geographical location, time of year, and time of day.
Included in these calculations is the atmospheric composition based on 3-hourly CAMS analyses (aerosols, water vapour, and various other gases).
Accuracy of global horizontal irradiance is within several percent \citep{gschwind_improving_2019}, enough for our purposes in this study as we will look at measurements of significantly larger variations.
It is available at a 1 minute resolution, which we linearly interpolate to 1 second when necessary.

For both context and independent validation of measured spectral irradiance, we calculate clear-sky shortwave irradiance spectra using libRadtran \citep{emde_libradtran_2016}.
Aerosols are set to default (rural-type boundary layer aerosol), total column water vapour is taken from microwave radiometer (FESSTVaL) or sounding (LIAISE) measurements, surface albedo is set to "cropland", and other atmospheric profiles are set to the "mid-latitude summer" default.
Other relevant settings are the coordinates and time of day, which are case-specific.
Validation of the setup is done using four clear-sky moments of the FESSTVaL campaign: 8:00 UTC on June 17, 18, 27, and 11:30 UTC on June 18. 
Values of clear-sky irradiance overestimate the Falkenberg sun tracker observations by 0.8 to 1.0 \% (5.5 - 7 W m$^{-2}$) for the 8:00 UTC cases, and by 1.3 \% (11 W m$^{-2}$) for 11:30 UTC on June 18. 
Performance for diffuse irradiance is significantly worse, likely due to using prescribed default aerosols, and is overestimated in all cases between 25 and 34 \%, except for 8:00 UTC June 17 with 9 \%. 
The impact of this bias is small in the context of this study, but will be taken into account when results are discussed. 
An overview of all simulations with validation statistics is available in Table S1.
The spectrum of June 18 at 11:30 UTC is illustrated in Figure \ref{fig:frostssi}.

Finally, for sensor cosine response corrections and calibration, we need accurate solar zenith and azimuth angles, which we calculate using PySolar \citep{pysolar}.

\section{Spatial patterns of surface solar irradiance}\label{sec:broadband}

\subsection{Deriving global horizontal irradiance}\label{sec:bbcalib}
Broadband solar irradiance is the total shortwave surface solar irradiance, often called global horizontal irradiance (GHI).
GHI is measured using pyranometers, or in combination with pyrheliometers (Figure \ref{fig:scheme_rad}b), and typically has a spectral range encompassing most or all of the shortwave irradiance spectrum (e.g. 200 to 3600 nm for the CMP22, \cite{kipp-cm22}).
This range exceeds the spectral range covered by the 18 wavelength bands of our sensor (410 to 940 nm, a bit more including the crosstalk sensitivity), but these bands cover the most energetic part of the spectrum (410 to 940 nm $\approx$ 68 \% of total irradiance based on data in Figure \ref{fig:frostssi}).
We derive the GHI by first taking the mean of all spectral bands (Figure \ref{fig:frostssi}), correcting for the cosine response, and converting raw measurement units, counts bin$^{-1}$ $\Delta$t$^{-1}$, to W m$^{-2}$.
This calibration generalises under the assumption that the spectrum shape of short-wave irradiance remains constant.

\begin{figure}[ht]
    \centering
    \includegraphics[width=\textwidth]{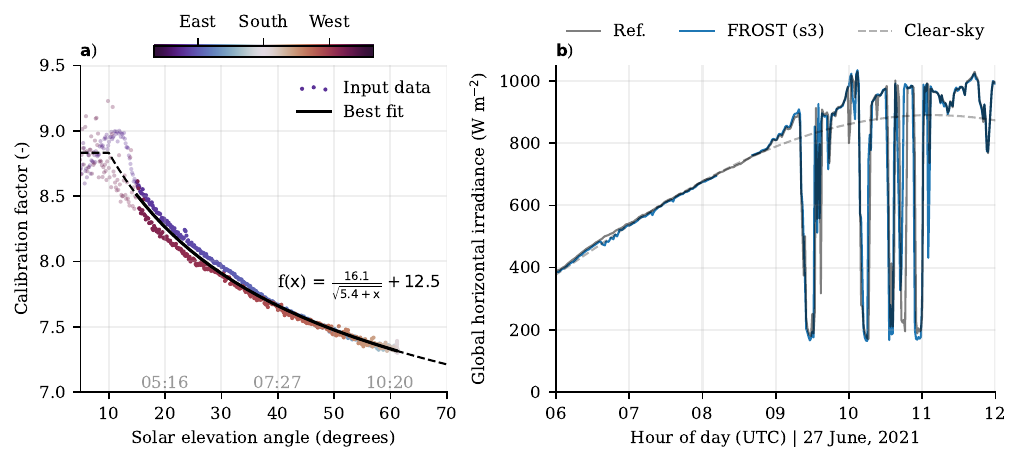}
    \caption{\textbf{Broadband global horizontal irradiance calibration example.} Calibration factor as function of solar elevation angle between raw instrument data and the irradiance data from the Falkenberg suntracker for June 18 is shown in \textbf{(a)}. The best fit curve is manually extended at the lower and upper end of the data range (dashed line), and follows the function with best fit coefficients between 15 and 60 degrees elevation angle. Time at the bottom is in UTC for morning until noon. The scatter colors indicate the solar azimuth angle. In \textbf{(b)}, the best fit calibration from \textbf{(a)} is applied to measurements of June 27, and compared again to the Falkenberg suntracker reference. All FROST data shown is for sensor 3, resampled to 1 minute.}
    \label{fig:bbcalib}
\end{figure}

For both FESSTVaL and LIAISE, we have at least one clear-sky day with high quality reference GHI measurements available from pyranometers to calibrate against.
The cosine response is a function of solar zenith angle, so we take the ratio between the spectral average of our instrument and the reference GHI measurement as a function of this angle.
Figure \ref{fig:bbcalib}a shows the resulting ratio curve relative to the suntracker reference for a single radiometer at FESSTVaL for a clear-sky day (June 18, 2021).
The best curve fit extrapolated from 15 to 10 degrees, and kept constant for any value below, because here the absolute signal gets too low and relative measurement errors, including that of the pyranometer cosine response, negatively impact the curve fit.
Especially differences between the cosine response of each subgroup become more pronounced at such low angles.
Furthermore, when the sensor is not perfectly level, measured incoming irradiance between morning and evening for the same solar elevation angle introduces an asymmetry, which explains part of the hysteresis effect visible in Figure \ref{fig:bbcalib}a.
We find that a fitting technique that takes the solar azimuth angle into account and fits on subgroups separately does not generalise beyond the calibration data, and thus does not improve data accuracy for days with clouds.
This is likely due to slight variations (less than 0.5 degrees) in sensor orientation that may occur from day to day, the exact reasons of which we do not know, and different combinations of solar elevation and azimuth angle as the Earth's orbit around the Sun progresses.
Figure \ref{fig:bbcalib}b illustrates the resulting time series when the reference date best fit is applied to a different day (June 27, 2021).
This examples shows sensor 3 captures both the daily cycle and cloud driven fluctuations.

\begin{figure}[ht]
    \centering
    \includegraphics[width=\textwidth]{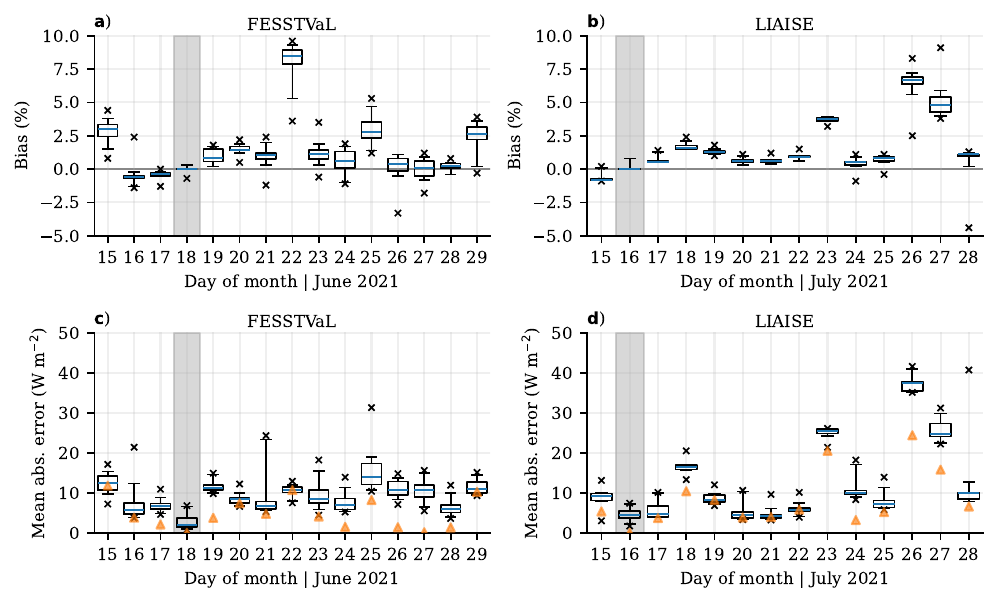}
    \caption{\textbf{Global horizontal irradiance calibration performance of all sensors for FESSTVaL and LIAISE}. Panels \textbf{(a, b)} show the mean bias of daily sums of irradiance compared to a reference pyranometer for all sensors, and panels \textbf{(c, d)} the mean absolute error. Vertical shading indicates the reference clear-sky calibration date for FESSTVaL in \textbf{(a, c)} and LIAISE in \textbf{(b, d)}. Box plots span the 5-95th percentile range of the spread among sensors. The orange triangle markers in \textbf{(b, d)} indicate the absolute median bias in W m$^{-2}$. Only data with solar elevation angle above 15 degrees is used for verification.}
    \label{fig:bbcalib_overview}
\end{figure}

We apply this calibration routine to all data for each sensor individually, for both campaigns separately.
Figures \ref{fig:bbcalib_overview} give an overview of the performance of all grid sensors across all campaign days for FESSTVaL and LIAISE, compared to their respective reference station. 
Overall, the bias is within 2\%, and mean absolute error (MAE) between 5 to 15 W m$^{-2}$ for most days and sensors. 
Notable outliers, June 22 or July 26 and 27, are explained by overcast and rainy weather (Figure \ref{fig:campaignwx}), which leads to increased absorption in mostly near-infrared wavelength bands that our instrument does not sample, but is part of the shortwave spectrum.
GHI is thus overestimated, because clear-sky conditions with less relative absorption in the near-infrared are the basis for calibration, violating our assumption that the solar spectrum shape is constant under all conditions (see \cite{heusinkveld_new_2023}, their Section 3.3, for more details).
Sensor to sensor variations are typically smaller or equal to the error with respect to the reference station, and likely originate from variations in construction and minor day to day changes in sensor orientation. 
Changes in calibration quality due to the sensors' temperature sensitivity are small, as for most days the measured 10 cm temperature, and therefore the Teflon diffuser temperature, was well above the 2\% signal jump at 20 degrees Celsius (Figure \ref{fig:campaignwx}).

\subsection{First impression and video}\label{sec:video}
\begin{figure}[ht]
    \centering
    \includegraphics[width=\textwidth]{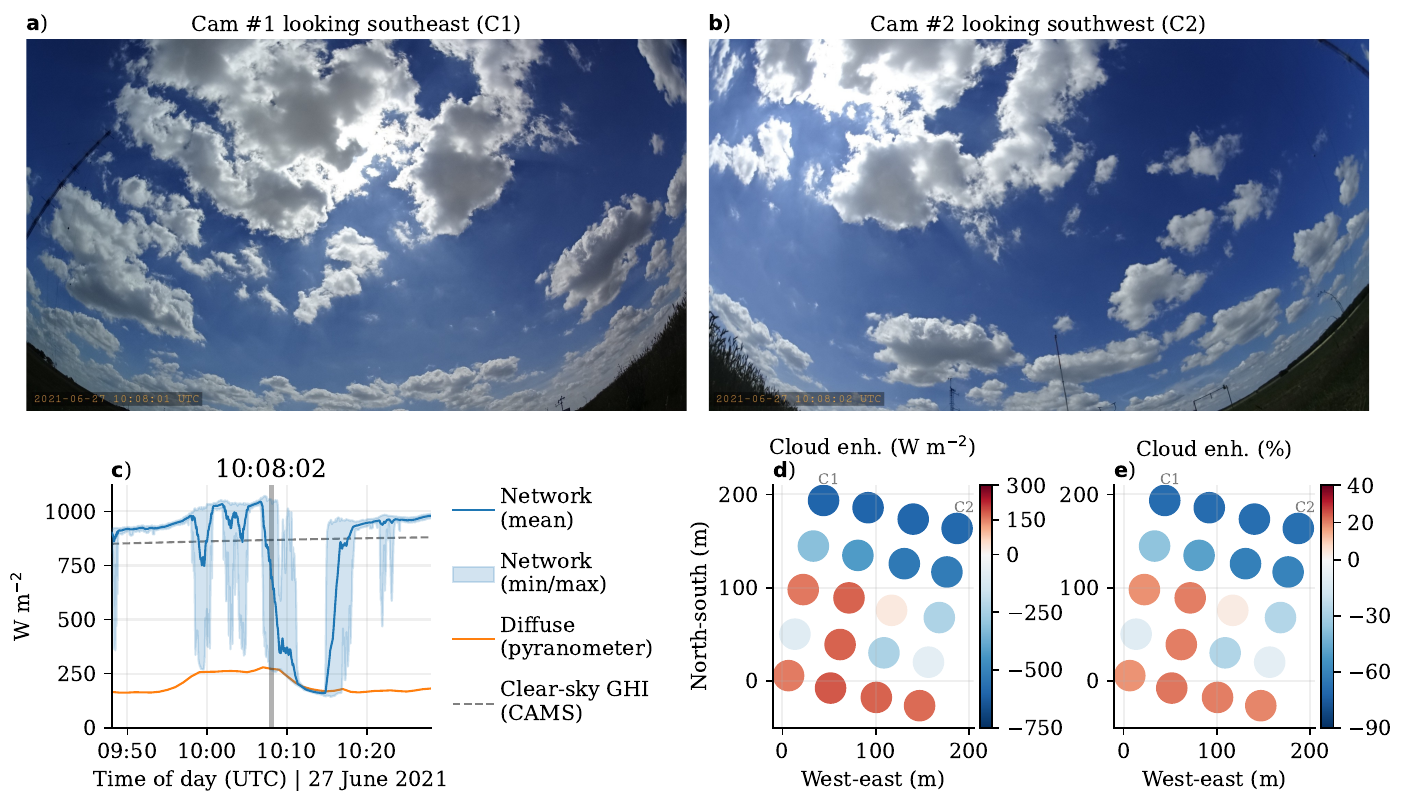}
    \caption{\textbf{Network measurements of GHI combined with cloud imagery} for June 27, 2021, at Falkenberg (FESSTVaL campaign). The time series \textbf{(c)} are centered around the 11:00 UTC snapshot, and features diffuse irradiance from the Falkenberg sun tracker, clear-sky GHI, and the FROST network measurements (spatial mean and min/max range). The network measurements are plotted relative to clear-sky GHI (CAMS McClear) in an absolute \textbf{(d)} and relative way \textbf{(e)}. For an animated version, see supplementary material or \url{https://vimeo.com/827602111}.}
    \label{fig:cloudcamframe}
\end{figure}

To get a first impression of cloud-driven patterns of surface solar irradiance, we plot the sensor network directly on a map combined with the cloud camera images, and render frames for each time step to create a video.
For June 27, 2021, a day with cumulus clouds at FESSTVaL, such a video is available on \url{https://vimeo.com/827602111} (or supplementary material).
One frame is displayed here in Figure \ref{fig:cloudcamframe}, but in particular the video shows how the sensor network captures irradiance patterns made by the dynamic cumulus field.
Some interesting features are how some cloud passages show a clear temporary increase in diffuse irradiance, whereas for others the diffuse irradiance remains constant, albeit above clear-sky irradiance (Figure \ref{fig:cloudcamframe}c).
There is an extra step of complexity once cirrus fields pass over, superimposing their effect onto that of the boundary layer cumuli.
It is also clear that the spatial scale of cumulus shadows and enhancements is well above that of the network size and transitions between shaded and sunlit areas occur at scales smaller than the network sensor spacing.  

\begin{figure}[ht]
    \centering
    \includegraphics[width=\textwidth]{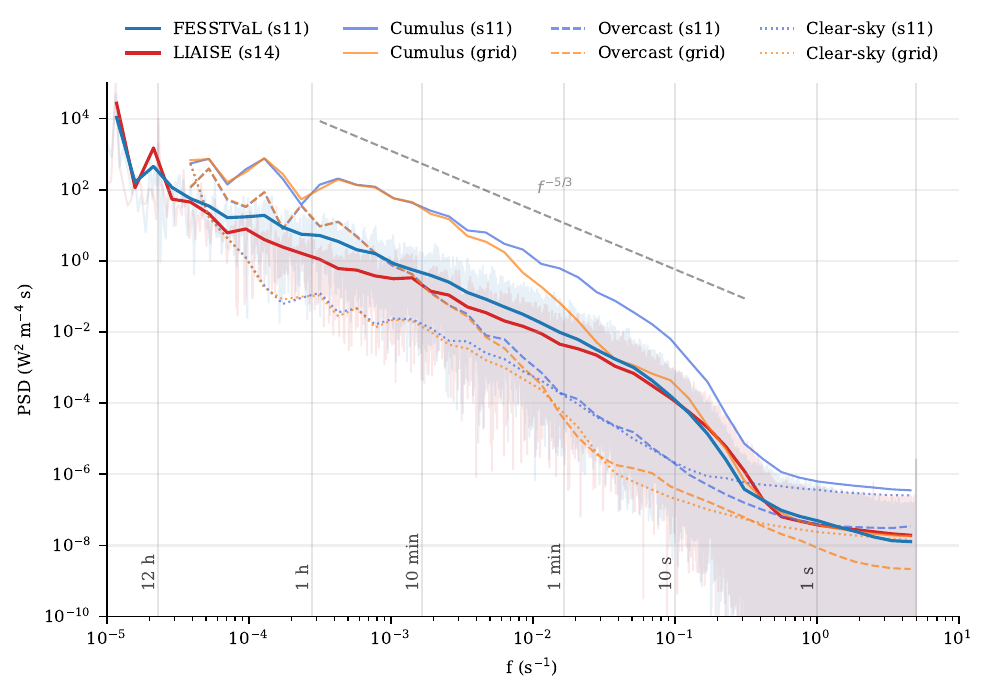}
    \caption{\textbf{Power density spectra for global horizontal irradiance time series} of the whole FESSTVaL and LIAISE campaigns. A cumulus (June 27), overcast (June 22), and clear-sky day (June 18) are illustrated separately for both a single sensor and the average of the Falkenberg grid.}
    \label{fig:psd_all}
\end{figure}

As an illustration of overall observed variability throughout both field campaigns, we construct a power density spectrum of all time series in FESSTVaL, LIAISE, and for three cases separately, illustrated in Figure \ref{fig:psd_all}.
The campaign averaged spectra follow approximately f$^{-5/3}$ scaling between 10$^{-4}$ and 10$^{-1}$ s$^{-1}$, before a scale brake between 10$^{-1}$ and 10$^{0}$ and an apparent continuation of weaker power law scaling thereafter.
Similar f$^{-5/3}$ scaling is shown by \cite{tabar_kolmogorov_2014} until $\sim$ 10$^{-1}$ s$^{-1}$, after which it similarly deviates. 
The scale break between 10$^{-1}$ and 10$^{0}$ s$^{-1}$ is expected, at least for broken cloud conditions, due to the smallest clouds becoming transparent to solar irradiance \citep{mol_reconciling_2023}, making the biggest source of variability disappear.
Other studies using spatial pyranometer networks find clearly different spatiotemporal scales and magnitude of variability for different sky types, and identify 'broken clouds' as the most potent for generating variability at the smallest scales \citep{lohmann_local_2016, madhavan_multiresolution_2017}.
Power density spectra for single sensor time series of clear-sky (June 18), overcast (June 22), and cumulus (June 27) show differences in power law scaling in the 10 to 3600 s range, the variance in cumulus dominating over other sky types across all scales, and overcast conditions having the lowest variance of all sky types at scales shorter than 10 s, all consistent with spatial wavelet variance presented in \cite{madhavan_multiresolution_2017} (their Figure 5).
Grid-averaged spectra start to deviate from single point measurements at scales shorter than 10 minutes.
The spectral power of Falkenberg grid average is an order of magnitude below that of a single sensor at these small scales, relatively consistent between the three sky types. 

\subsection{Three types of patterns}
\begin{figure}[ht]
    \centering
    \includegraphics[width=\textwidth]{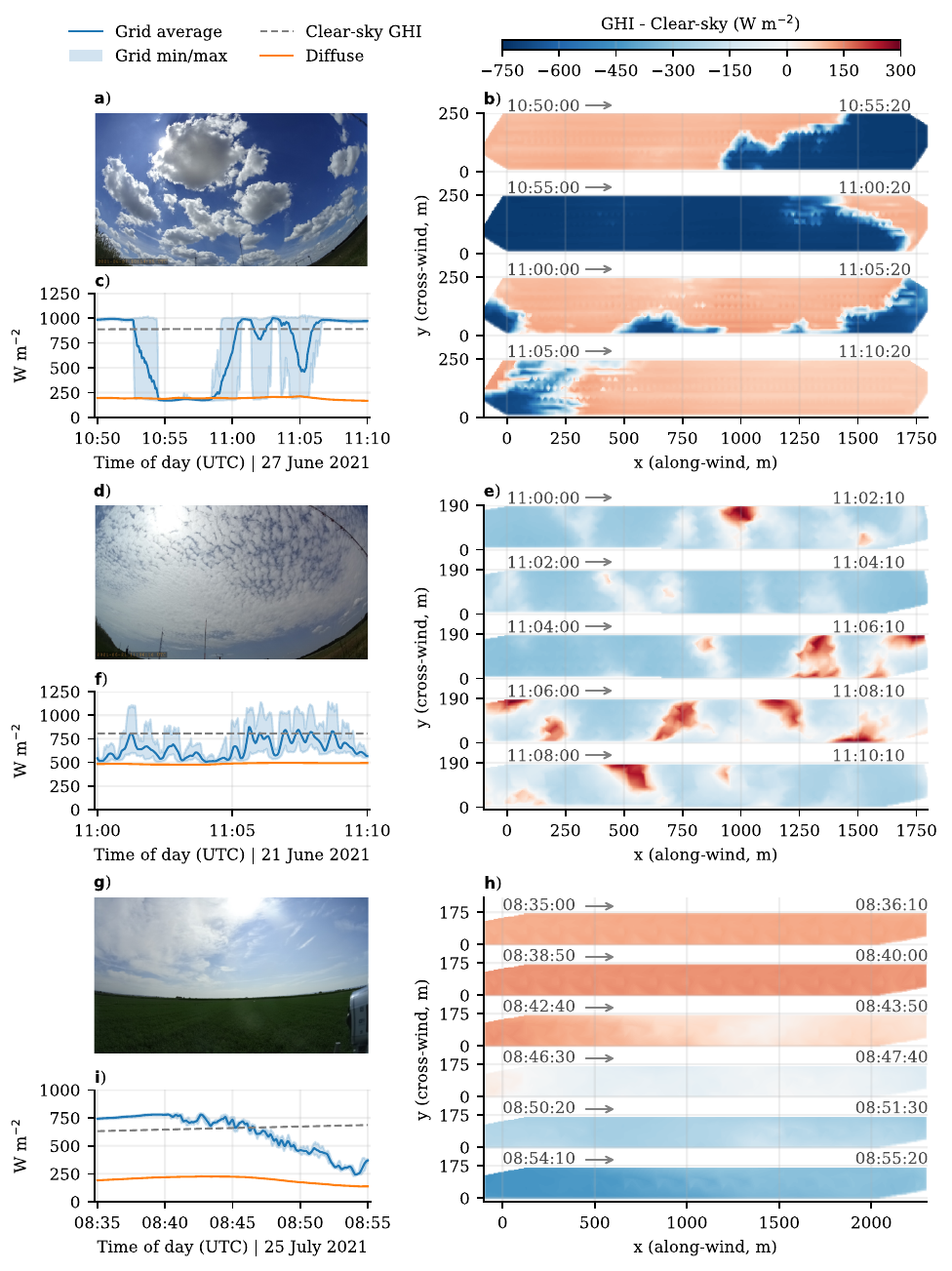}
    \caption{\textbf{Spatial patterns of cloud enhancements and shadows} for three cases: cumulus \textbf{(a-c)}, altocumulus \textbf{(d-f)}, and cirrus \textbf{(g-i)}. \textbf{(a, d, g)} show a representative photo. The time series include grid measurements (FROST network at $\Delta$t = 1 s), clear-sky GHI (CAMS McClear at $\Delta$t = 60 s), and diffuse irradiance (Falkenberg sun tracker \textbf{(c, f)} at $\Delta$t = 60 s or La Cendrosa energy balance station at $\Delta$t = 5 s \textbf{(i)}). Data in \textbf{(b, e, h)} are plotted relative to clear-sky irradiance. The temporal range of each pattern segment is labelled in HH:MM:SS on top.}
    \label{fig:ghi_patterns}
\end{figure}

Often during the campaign, a combination of cloud types are present with varying degrees of total cloud cover and optical thickness, for example a mix of cumulus (passive to deep convective), multiple layers of altocumulus, and/or cirrus.
To demonstrate the diversity of observed surface irradiance variability during the campaigns, we select three cases.
These selected cases are of cloud types that occur in isolation, making the interpretation easier, and are frequently present in the campaign period: 1) boundary level cumulus, 2) mid level altocumulus, and 3) high level cirrus, illustrated in Figure \ref{fig:ghi_patterns}.
The patterns are visualised using the technique explained in Section \ref{sec:spatialinterp}, and are robust to sensor exclusion tests (shown in Figures S4 and S5, for the first two cases).

Case 1 is from the same date as Figure \ref{fig:cloudcamframe} (June 27, 2021), and features slow-moving (4.5 $\pm $ 1 m s$^{-1}$) fair weather cumuli with cloud bases at $\sim$ 1750 m, and a relatively clean atmosphere (deep blue sky).
Cloud enhancement to shadow transitions are very sharp, about 40 to 60 m as estimated from Figure \ref{fig:ghi_patterns}b, making cumulus cloud shadows slightly smaller than their size with a length scale similar to what we found in \cite{mol_reconciling_2023} based on long-term time series.
Diffuse irradiance does not vary significantly on the spatiotemporal scales of this example, suggesting the total light scattering off of clouds in this case has a very wide horizontal range without much contribution to the total enhancement from local forward scattering at transparent cloud edges.
Around the smaller cloud fragments (11:00 to 11:05 UTC, bottom pattern in Figure \ref{fig:ghi_patterns}b), we do observe an increases in cloud enhancement of $\sim$ 30 to 40 W m$^{-2}$ (4 \% of clear-sky), and overall the diffuse is lower after the passage (11:10 UTC) than before (10:50 UTC).
We think the absence of diffuse peaks close to transitions in this case may be due to the relatively high cloud base, where forward scattering through cloud edges is spread out over a large surface area rather than more locally focused.
However, the suntracker measurements are 1 minute averages, so it may miss local peaks in diffuse irradiance, and changes in the sources of scattered light (overall cloud field vs. local cloud edge) could coincidentally be counteracting.
The contrast between cloud enhancement and shadow is large, approximately 750 W m$^{-2}$, or 80 \% of clear-sky irradiance.
Some of the artefacts arise from small biases between sensors, such as the horizontal stripes at 10:50 and 11:10 UTC, or likely changing cloud shape, such as the noisy pattern around 11:06 UTC. 
Section \ref{sec:spectral} introduces a few more cumulus passages of this case, including spectral effects, to further investigate what is going on.

The second case (Figure \ref{fig:ghi_patterns}d) is of an altocumulus field at 5 km altitude (estimated using the ceilometer of Falkenberg), moving at 14.3 m s$^{-1}$, and under overall hazier conditions than the first case.
Individual altocumulus clouds are about 1 to 5 times the area of the sensor network based on the distance between cloud enhancement peaks, whereas these peaks are up to $\sim$ 200 m in diameter and thus mostly fit within the network area.
Apart from the spatial scales being significantly smaller than the cumulus case, the shadow patterns are weak (250 W m$^{-2}$ or 35 \% below clear-sky), and cloud enhancements very strong, locally more than 300 W m$^{-2}$ or 40 \% above clear-sky.
The mechanism appears to be a consistently high diffuse irradiance (500 W m$^{-2}$, two-thirds of clear-sky) generated by forward scattering in the semi-transparent altocumulus field as a whole, with superimposed gaps in the clouds that let up to 650 W m$^{-2}$ of direct irradiance locally pass through.
These are thus particularly extreme variations at small spatiotemporal scales, with a similar relative magnitude (cloud shadow to enhancement contrast) to that of the previous example with cumulus clouds.

The third and last case (Figure \ref{fig:ghi_patterns}g) is features an optically thick cirrus field moving over the LIAISE network setup at 35 m s$^{-1}$ (July 25, 2021), and is notably different from the other two cases. 
Only weak spatial patterns of about 500 meters in length are visible in some parts of the cirrus (8:43 and 8:46 UTC for example, Figure \ref{fig:ghi_patterns}h).
Despite the high cloud velocity, it takes 15 minutes to go from a 120 W m$^{-2}$ or 18\% of cloud enhancement (8:39 UTC) to a near complete blocking of direct irradiance (8:54 UTC). 
The area of influence of this patch of cirrus far exceeds the radiometer grid size, with the transition from cloud enhancement to shadow minimum covering 35 m s$^{-1}$ $\times$ 900 s = 31.5 km.

The scales of patterns these three cloud types make suggest errors made by radiative transfer models using the independent column approximation (no 3D effects) will already become apparent for cirrus in coarser resolution medium range numerical weather prediction (e.g., ECMWF's IFS at $\sim$ 9$\times$9 km$^{2}$), not just at the smaller scales of broken boundary layer clouds in cloud resolving models.
These three cases demonstrate that the spread within the category of 'broken cloud', here extended beyond boundary layer clouds, is large, and furthermore that not just cumulus, but also mid and high level clouds are worth further attention given their common occurrence globally (e.g. \cite{sassen_classifying_2008}) and large effects on surface irradiance.

\clearpage

\section{Spectral signature of cloud-driven irradiance patterns}\label{sec:spectral}
In this section, we expand upon the three cases by looking at changes in spectral irradiance defined as a ratio between short and long visible wavelengths.
The motivation and calculation of this ratio is described first. 

\subsection{Identifying changes in 'blue' versus 'red' light}
Clear-sky conditions are famously characterised by blue skies: diffuse irradiance enriched in shorter wavelengths and a yellow sun due to depletion of those same wavelengths, owing to the $\lambda^{-4}$ dependency of Rayleigh scattering.
An example of the spectral signature of the total irradiance and the diffuse component is illustrated in Figure \ref{fig:frostssi}, illustrative for a mid-latitude summer with standard rural aerosols (e.g. no wildfires or Sahara dust). 
As clouds interact with both components of irradiance, we expect to see changes in the ratio between short and long wavelengths as the relative contribution of the scattered and non-scattered light to the total surface irradiance varies. 
Such an analysis requires scattering and absorbance of light by clouds to have no significant dependency on  wavelength, true for the most energetic part of the solar spectrum \citep{ackerman_absorption_1987, ohirok_three-dimensional_1998-1, key_parameterization_2002, schmidt_apparent_2010}, and all other conditions kept equal (aerosol amount, spectral surface albedo, gas concentrations).
For example, if a single cumulus cloud in an otherwise blue sky blocks direct (blue-depleted) irradiance, we expect the spectrum of remaining light to be relatively enriched in short wavelengths.
Hygroscopic growth of aerosols in air with high relatively humidity, often found near cloud edges, causes aerosols to scatter more, and more so for shorter wavelengths.
This was found by \cite{gristey_influence_2022} in cases of shallow cumuli to be an important contribution to enhanced diffuse irradiance in cloud shadows.
Despite constant aerosol amount, this effect may also further deplete the direct irradiance near cloud edges of shorter wavelengths, thereby contributing to more 'blue' cloud shadows. 

In this section we explore spectral effects of the same cases as demonstrated in Figure \ref{fig:ghi_patterns} by comparing the ratio between the shorter and longer wavelengths of the part of the irradiance spectrum that contains most energy. 
We choose multiple bands of the radiometer to get a stronger signal, and further limit our choice by excluding bands of $\lambda < $ 485 nm (too high crosstalk, Section \ref{sec:crosstalk}) and $\lambda > $ 700 nm (growing contribution of spectral effects of water).
Specifically, we choose the bands $\lambda_s$ = (485, 510, 530 nm) and $\lambda_l$ = (645, 680, 705 nm) and define the ratio $r = \lambda_s / \lambda_l$ to represent 'blue' versus 'red' light.

Under a flat spectrum, the crosstalk for the chosen bands are $ct_s \approx$ 45 \%, $ct_l \approx$ 14 \% for $\lambda_s$, $\lambda_l$, respectively.
Crosstalk is roughly halved for a more typical clear-sky spectrum (as in Figure \ref{fig:frostssi}).
We cannot a priori correct for crosstalk, because it would require knowing the spectrum of irradiance and how it changes, whereas this is what we want to measure and characterise.
However, we can simulate how changes in measured $r$, $\Delta r_m$, relate to true changes $\Delta r_t$, building upon our assumption that nothing influences $r$ but changes in the mixing of clear-sky diffuse and direct irradiance.
To validate the approach and interpret results, we supplement the analysis with simulated clear-sky spectra (Section \ref{sec:modelinfo}) for each case, which also provide independent estimates of $r_{cs}$, the clear-sky ratio.
Taking the clear-sky diffuse illustrated in Figure \ref{fig:frostssi} and mixing in direct irradiance from 0 to 1.6 the clear-sky value, we find a linear relationship $\Delta r_m$ = $\gamma \Delta r_t$ and $\gamma \approx$ 0.5, i.e., measured changes are underestimated due to crosstalk (Figure S6).
Both \cite{durand_diffuse_2021} (their Figure 2) and a simulation with a homogeneous water cloud ('fval\_j18\_wc', Table S1) show that $r$ would decrease in overcast conditions compared to clear-sky, but that changes in the $\lambda_{ct}$ range are insignificant.
Furthermore, \cite{heusinkveld_new_2023} show that the sensor version without crosstalk has qualitatively the same effect of overcast compared to clear conditions (their Figure 17).
We are therefore confident that the sign of change we observe is correct and that the magnitude of change is underestimated. 
Absolute values will nonetheless be interpreted with caution.
Because the measured spectral data are in raw sensors units, we apply a one-time calibration factor $r = 1.4 r_m$ based on comparison of four simulated and measured clear-sky spectra.

\subsection{A time series of spectral changes}
\begin{figure}[ht]
    \centering
    \includegraphics[width=\textwidth]{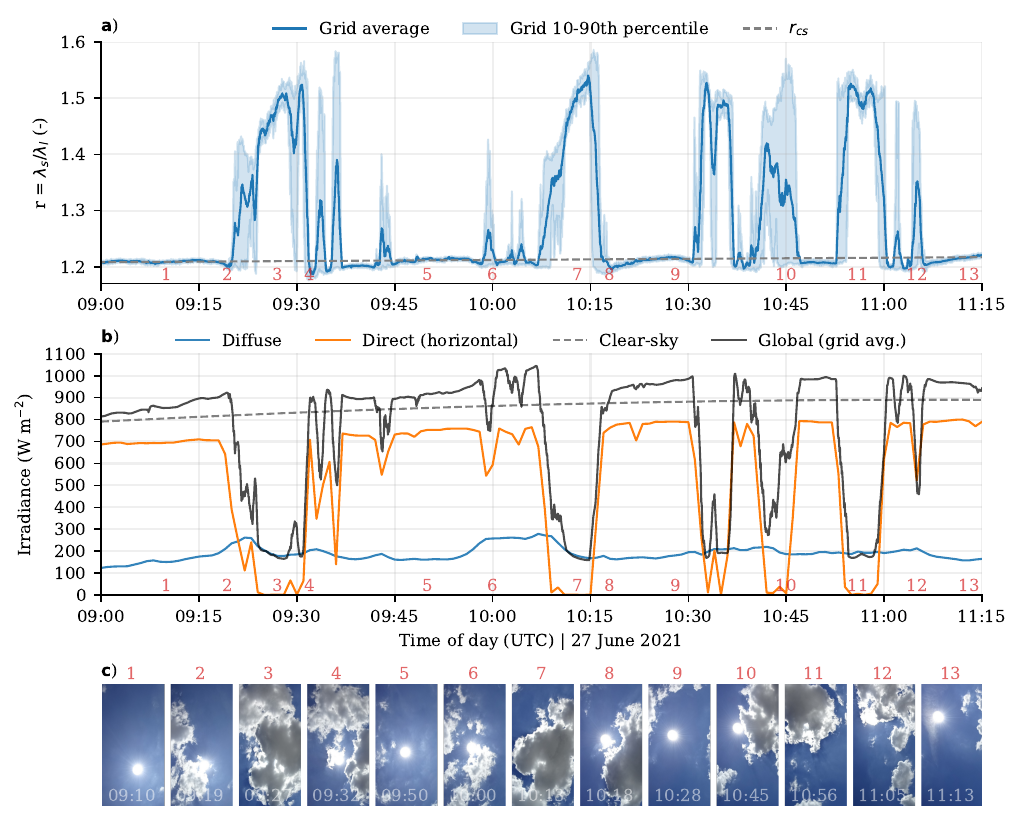}
    \caption{\textbf{Time series of $r$ as measured by the network at FESSTVaL} on 27 June 2021 in \textbf{(a)}, combined with broadband irradiance to illustrate the total effect of cloud passages in \textbf{(b)}. Direct and diffuse irradiance are 1-minute averages of the Falkenberg suntracker, clear-sky is based on CAMS McClear. Global horizontal irradiance is 1-Hz data based on the mean values in the FROST network at Falkenberg. In \textbf{(c)}, cropped images of one of the cloud cameras are shown for 13 manually chosen points in the time series to illustrate degrees of direct light obstruction of cumulus clouds.}
    \label{fig:spectrum_ts}
\end{figure}

Figure \ref{fig:spectrum_ts} shows how $r$ varies in relation to multiple shading and enhancement events of cumulus clouds on June 27 (FESSTVaL). 
The time series underlines that the fairly constant diffuse irradiance seen in the first case of cumulus is not representative of all cumulus passages, as most others show local distinctly enhanced diffuse irradiance.
As for $r$, most noticeable are the large shifts ($\Delta r_m \approx$ 0.3) towards blue-enriched light in fully shaded conditions, e.g. at 09:26 or 10:13 UTC. 
Further away from transitions, e.g. 09:10 and 09:50 UTC, there is enhanced irradiance, but no clear deviations from estimated clear-sky ratios.
Interestingly, every shading event is flanked by brief reductions in $r$, typically 0.01-0.02 below $r_{cs}$, for example at 09:19 and 09:32 UTC, just as the sun illuminates the local cloud edge as seen from the surface.
Furthermore, it appears not every cloud enhancement has the same proportions of $\lambda_s$ and $\lambda_l$, as there are different combinations of cloud enhancement magnitude and $\Delta r_m$ (e.g. 09:19 vs 09:32 UTC).
This puts some value to the idea that the lack of extra local irradiance enhancement around shadows observed in the first case of Section \ref{sec:broadband} is due to counteracting local and non-local scattering sources.
We speculate that the horizontal distribution of light scattered off (or escaping from, \cite{varnai_effects_1999}) cloud sides is approximately a clear-sky mix of diffuse and direct light ($r = r_{cs}$, $\Delta r_m = $ 0) and acts over a larger area further away from the cloud(s), whereas localised forward scattering right at cloud edges is mostly direct light with lower $r$ ($\Delta r_m \approx$ -0.01 to -0.02).
We will take a closer look at this phenomenon spatially using the cumulus, altocumulus, and cirrus cases.

\subsection{Spatial patterns of spectral changes}
\begin{figure}[ht]
    \centering
    \includegraphics[width=\textwidth]{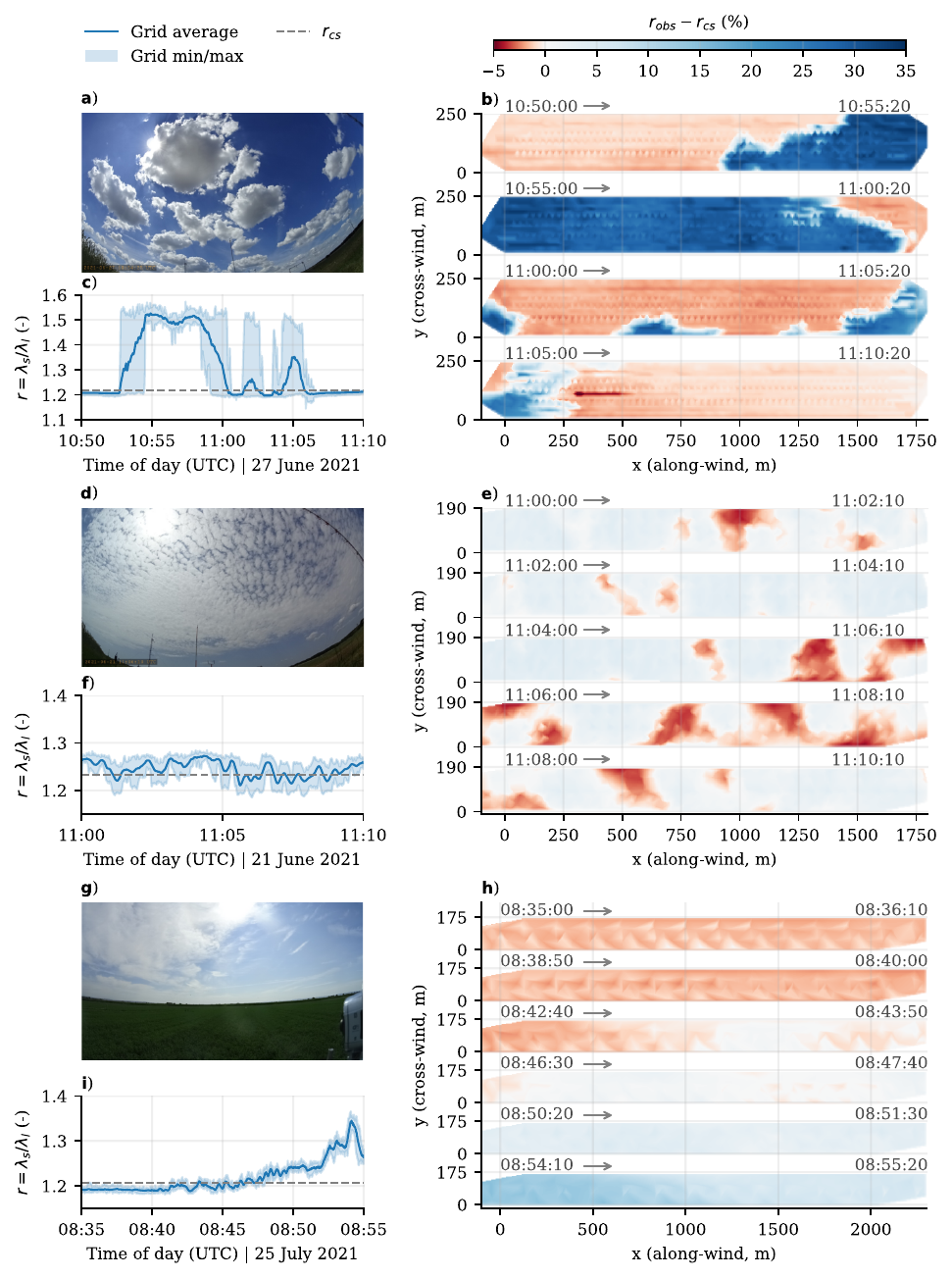}
    \caption{\textbf{Spatial patterns of changes in $r$ compared to clear-sky}. Lower values indicate red-enrichment, higher values blue-enrichment. Values for $r_{cs}$ are based on simulated clear-sky spectra. Note the asymmetric colormap scaling. Similar layout as Figure \ref{fig:ghi_patterns}}
    \label{fig:r_patterns}
\end{figure}

Figure \ref{fig:r_patterns} shows $\Delta r_m$, now expressed in percent-points, for the cumulus, altocumulus, and cirrus case, previously described in Section \ref{sec:broadband}. 
For each case, data is plotted relative to the clear-sky ratio estimated using a simulated spectrum, resulting in another diverse set of patterns, albeit more noisy than for broadband irradiance due to reduced signal strength.

We first zoom in on four areas in the cumulus case (Figure \ref{fig:r_patterns}a-c): 10:51, 10:56, 11:01, and 11:10 UTC.
For the three enhanced areas, the shift in the middle one (11:01 UTC) is about twice as strong ($\Delta r_m$ = -1.8 \%) compared to before (10:51 UTC, $\Delta r_m$ = -1.1 \%) and after (11:10 UTC, $\Delta r_m$ = -0.8 \%), whereas the cloud enhancement in broadband irradiance only varies between 9 (11:10 UTC) and 12 \% (10:51 UTC).
In the shadow at 10:56 UTC, $\Delta r_m$ = 30.4 \%.
If we assume all the light in the shadow comes from clear-sky diffuse, $\Delta r$ would be 74.7 \%.
Adding clear-sky direct irradiance to match the observed diffuse irradiance gives 60.0 \%, or 45.0 \% if we add more direct irradiance to compensate for overestimation of clear-sky diffuse of $\sim$ 25 \% of the modelled spectrum at 08:00 UTC.
Here, we are making various assumption about the origin and properties of irradiance that are unrealistic, but they serve as an hypothetical situation of how $r$ would change, and how much of this irradiance 'mixing' would be necessary to induce changes in $r$ similar to what is observed.
The first and last approximation are under and over-estimations of the amount of blue-enriched irradiance in the shadow: the diffuse irradiance cannot be purely clear-sky diffuse given the present cloud field, nor would all enhanced diffuse light be of a direct irradiance origin.
The middle estimate may be correct for the wrong reasons, but it at least represents a case where a significant portion of irradiance is coming from horizontally scattered direct or total irradiance mixed in with clear-sky diffuse, and is of similar magnitude as $\Delta r_m$ if we assume the estimate $\gamma \approx$ 0.5 is valid.
Similarly, attributing all cloud enhancement to direct irradiance gives $\Delta r$ = -2.8 \% and -2.5 \% for 10:51 and 11:10 UTC, an overestimation even after a $\gamma$ correction, suggesting the extra irradiance is mix of spectral irradiance closer to that of clear-sky conditions.
For 11:01 UTC, however, the estimate of $\Delta r$ = -2.7 \% more closely matches the observed $\Delta r_m$ = -1.8 \% (or $\gamma \Delta r_m \approx$ -3.6 \%), and coincides with a higher degree of fragmented semi-transparent clouds that can effectively scatter direct irradiance forward.
Alternatively, optically thick cumulus may also reduce $r$ by blocking part of diffuse clear-sky irradiance, though we expect this to be of secondary importance and more non-local due to the approximately isotropic nature of diffuse irradiance.
Hygroscopic growth of aerosols in regions near cloud edges may also contribute to subtle changes in $r$ and make the regions near cloud edges more potent in forward scattering.   
Another cumulus case with notably lower cloud cover, higher cloud base, and high apparent haziness (June 17 at FESSTVaL, not shown), has qualitatively similar patterns around cloud shadows, but perturbations $\Delta r_m$ are significantly larger. 
An analysis beyond two cumulus case studies and controlling for cloud and aerosol optical properties is necessary before drawing more general conclusions. 

The altocumulus and cirrus cases are more tricky to analyse, as here the clouds are all semi-transparent and at higher altitude, thus making the origin of light and its spectral signature more complicated.
The relative spread in $r_m$ is also larger in both, for altocumulus because the patterns are similar to the network scale, and for the cirrus because it is earlier in the day with a weaker signal, visible in a noisier time series (Figure \ref{fig:r_patterns}f) or pattern (Figure \ref{fig:r_patterns}h).
Values for $\Delta r_m$ are nonetheless significant in both cases, and though the hypothetical mixing of clear-sky spectra will not help to identify the scattering mechanisms, they can put the numbers in context.
For the darkest shadow in the altocumulus case, at 11:04:15 UTC, $\Delta r_m$ = 3.8 \%, and assuming the increase of diffuse irradiance compared to clear-sky is all attributable to the clear-sky direct irradiance, $\Delta r$ = 8.8 \%, an expected overestimation (even after a factor $\gamma$) of blue-enrichment given that direct irradiance is $>$ 0 W m$^{-2}$.
For one of the stronger cloud enhancements (30.0 \% at 11:05:40), $\Delta r_m$ = -4.0 \%, and attributing all enhancement to direct irradiance also gives $\Delta r$ = -4.0 \%, likely an underestimation, suggesting part of the strongly enhanced diffuse irradiance may be more enriched in $\lambda_l$.
Averaged over the network, variations in $r$ are small ($|\Delta r_m| \approx$ 5 \%), about 6 to 7 times smaller compared to the cumulus case.
Lastly, for the cirrus case the enhancement is 20.9 \% at 08:39 UTC, with $\Delta r_m$ -1.5 \%, and $\Delta r$ = -4.6 \% assuming all enhancement is from clear-sky direct irradiance, a clear overestimate.
For the partially shaded area at 08:50 UTC we find $\Delta r_m$ = 3.1 \%, and attributing the increase in observed diffuse to clear-sky direct irradiance gives $\Delta r$ = 6.3 \% $\approx \gamma \Delta r_m$, except this ignores that over half of the observed light is direct irradiance (Figure \ref{fig:ghi_patterns}g).
Variations in $r$ here are in between the cumulus and altocumulus case ($|\Delta r_m| \approx$ 15 \%), but the spatiotemporal scale is two orders of magnitude larger. 

In summary, in all presented cases, both the cloud enhancement and shadow patterns show significant deviations from clear-sky spectral irradiance, which are particularly significant in magnitude and spatiotemporal scale for the cumulus clouds.
The fact various combinations of diffuse and direct irradiance can create cloud enhancements of the same magnitude is well known ( e.g. \cite{gueymard_cloud_2017}), and so is the fact clouds have spectral effects (e.g. \cite{ohirok_spectral_2000}). 
The spectral effect demonstrated in this section highlights there are changes in part of the spectrum due to differences in the origin of light, where otherwise the optical properties on the scale of these variations are mostly wavelength-independent.  
In particular the spectral signature of low and optically thick cumulus passages give some weight to our speculation that various light scattering mechanisms are at play, affecting different areas relative to the cloud.

\clearpage

\section{Temporal patterns driven by water vapour variability}\label{sec:wva}
One other significant spectral signature is that of water vapour.
In this section, we will look at two examples of variability in atmospheric moisture as measured by the FROST network at FESSTVaL and LIAISE.
These examples have a scope that is more towards regional scale (campaign area) land-atmosphere coupling and atmospheric dynamics under clear-sky conditions, rather than cloud-driven irradiance variability.
The variations in water vapour are not large enough to affect broadband irradiance in the way clouds do, but they may help explain observed variations in GHI in clear-sky conditions.

\subsection{Deriving total column water vapour}
\begin{figure}[htb]
    \centering
    \includegraphics[width=0.5\textwidth]{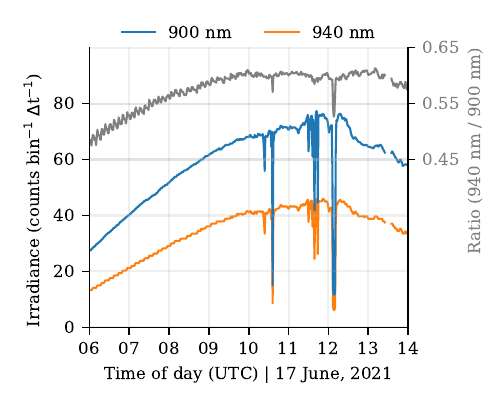}
    \caption{\textbf{Spectral signature of water vapour measured by the 900 and 940 nm bands of FROST}. The time series shows part of a daily cycle for a single sensor in the Falkenberg network, which is clear-sky until 10 UTC, but features brief cumulus passages thereafter. The ratio of the 940 nm / 900 nm bands is shown in grey on the right axis.}
    \label{fig:wva_example}
\end{figure}

Total column water vapour (TCWV) in the atmosphere is indirectly measurable using water vapour absorption bands, where heightened levels of water vapour result in significantly reduced signal within absorption bands compared to other wavelengths.
Figure \ref{fig:frostssi} highlights two of such bands for water, at 940 and 1130 nm.
A strong absorption band is captured by the 940 nm channel of FROST, which in theory means we are able to detect changes in atmospheric moisture when comparing the 940 nm signal to a reference band outside the absorption band.
Choosing a reference channel is limited by some sensor design and performance limitations.
The cosine response varies between the three subgroups, such that changes in the ratio between two channels (e.g. 940 vs. 860 nm) are partially a result of instrument imperfections (see also Section \ref{sec:frosterror}) rather than changes in absorption strength.
The choice of suitable channels is further limited by crosstalk at the shorter wavelengths of the subsensor with 940 nm, making the 900 and 940 nm channels the best options, despite 900 nm also partially being in a (weak) absorption band. 
Figure \ref{fig:wva_example} illustrates the signal at 900 and 940 nm, where early in the day the ratio under clear-sky conditions is lower (more absorption) than noon due to the longer path length of irradiance.
Shading from (semi-transparent) cumuli also gives distinct absorption signals, though seemingly only for the most optically thick cumulus passages in this example.
This might be due to the change in diffuse/direct partitioning, with a longer path length and thus absorption of diffuse irradiance, but any stronger statement requires a more careful analysis.
But this underlines why only clear-sky conditions are suitable for estimating TCWV, as we currently can not separate the effect of liquid water (or ice) from that of water vapour.
One more limitation is the signal strength for the individual bands being low compared to the signal of water vapour variations, even at high solar elevation angles.
For a clear signal we therefore take a moving average of 120 seconds or more, thereby effectively reducing the temporal resolution.
In a new version of FROST, the crosstalk and weak signal issues have been addressed \citep{heusinkveld_new_2023}.

\begin{figure}[ht]
    \centering
    \includegraphics[width=\textwidth]{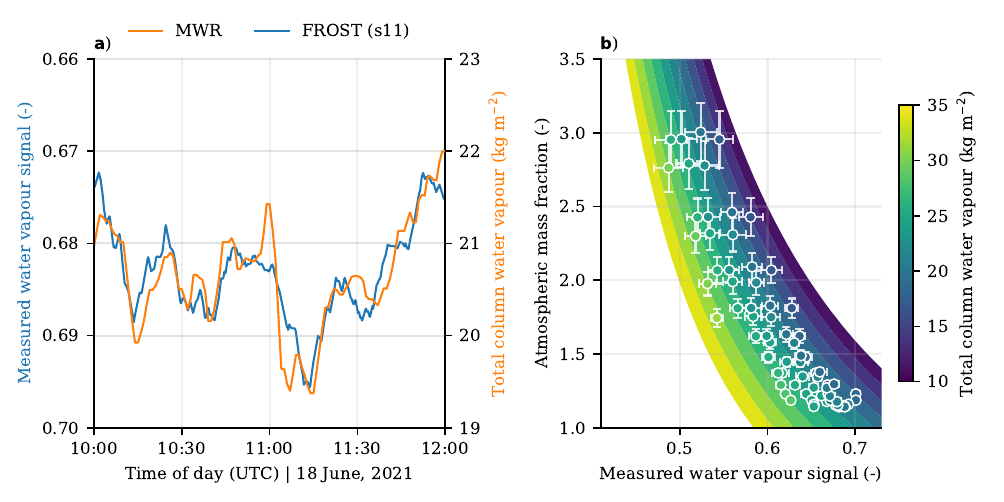}
    \caption{\textbf{Total column water vapour (TCWV) derived from the radiometer water vapour signal.} In \textbf{(a)}, an example time series of the signal (940 / 900 nm band ratio) compared to TCWV from the microwave radiometer \citep{ulrich_mwr_fesstal} is shown for June 18 at Falkenberg (FESSTVaL). In \textbf{(b)}, the scatters represent 30-minute averages and standard deviations of the measured water vapour signal and atmospheric mass fraction, coloured by the TCWV from the microwave radiometer. The results from the model best-fit are shown in the curved shading.}
    \label{fig:iwvfit}
\end{figure}

\begin{figure}[ht]
    \centering
    \includegraphics[width=\textwidth]{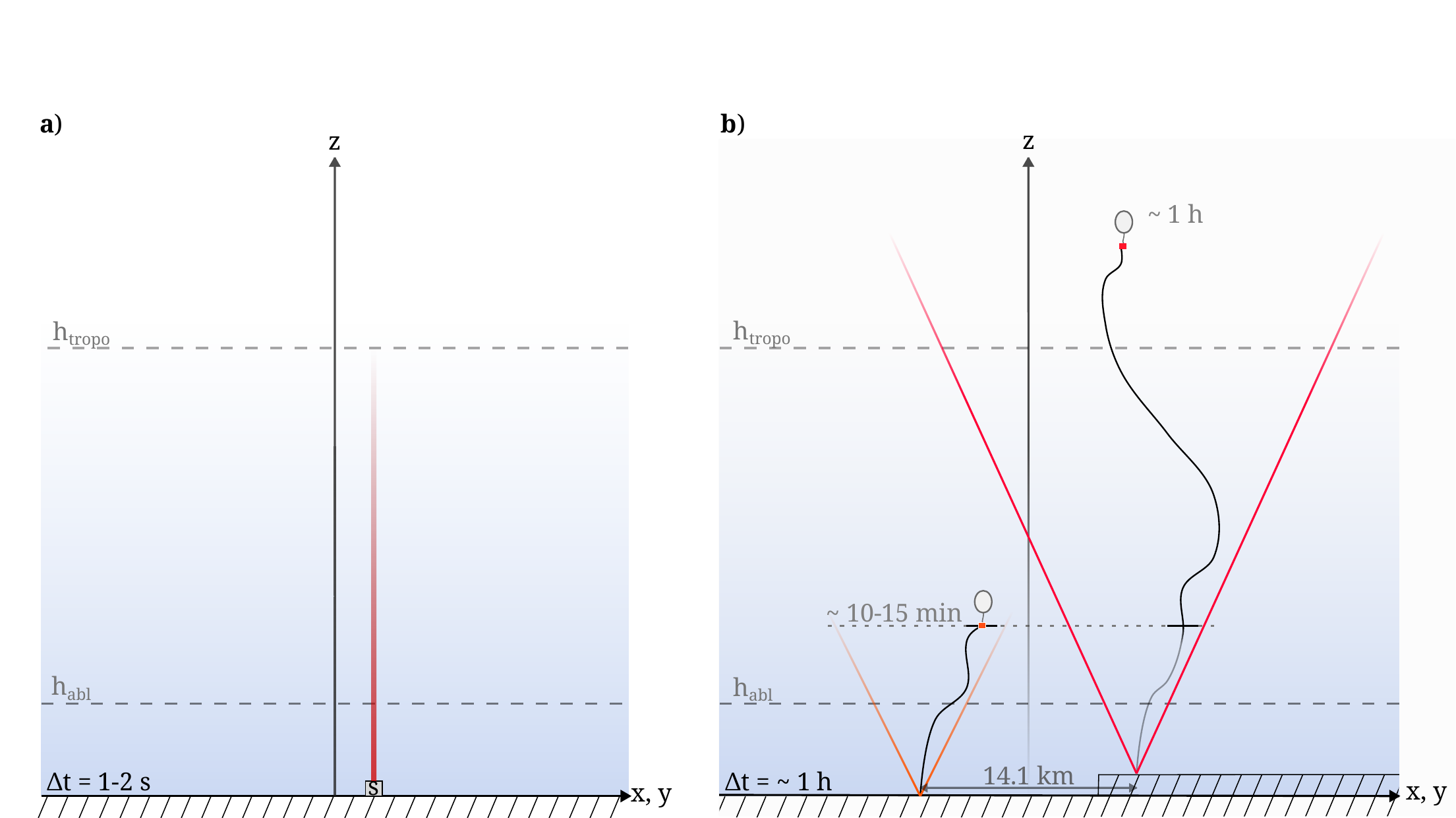}
    \caption{\textbf{Schematic of total column water vapour measurements} using a microwave radiometer (MWR) \textbf{(a)} and radiosondes \textbf{(b)}. The MWR is a passive instrument that measures humidity through microwave emissions from water vapour or liquid water, retrieved along a narrow vertical beam. To get the same measurement from soundings at La Cendrosa, shown in \textbf{(b)}, the boundary layer soundings (left, orange) are extended with tropospheric soundings (right, red) of Els Plans, 14.1 km apart and at slightly higher elevation. The colored cones illustrate the potential horizontal displacement with respect to the initial position, with the solid curved lines an example path a sounding could take. Measurement duration is approximately 10 to 15 minutes or 1 hour, respectively. The remainder of this plot is similar to Figure \ref{fig:scheme_rad}.}
    \label{fig:scheme_iwv}
\end{figure}

In Figure \ref{fig:iwvfit}a we show we can derive an absorption signal that is highly correlated to accurate reference measurements of TCWV.
Higher values in the 940 nm / 900 nm ratio indicate less TCWV, so by flipping the y-axis this correlation (r$^2$ = 0.75) is more clearly visible.
The reference TCWV measurement comes from a co-located microwave radiometer \citep{ulrich_mwr_fesstal}, with a 5 minute moving average applied to both time series to get a comparable signal. 
The microwave radiometer measures along a single straight vertical path (Figure \ref{fig:scheme_iwv}a), but the signal our sensors measure is a function of the path that light travels through the atmosphere, mostly from direct irradiance (Figure \ref{fig:scheme_rad}a). 
For fitting a model to the data we therefore include the atmospheric mass fraction (AMF), i.e. the path length light travels through the atmosphere given a solar elevation angle $\alpha$.
This also means that the light our instruments measure has a horizontal footprint of several kilometres, whereas the microwave radiometer is a vertical integral measurement at one fixed horizontal point.
Furthermore, under clear-sky conditions, diffuse irradiance still typically constitutes about 10 to 30\% of GHI (see Figure S1), and would have travelled a longer distance through the atmosphere (by definition, as it is scattered light compared to direct light). 
This effect is implicitly taken into account, as the diffuse fraction is also a function of solar elevation angle, and thus the atmospheric mass fraction.  
We expect some bias in our model fit for extremely clear or hazy days, but we have no clear signal or quantification of its effect within our observational dataset.

The relationship between TCWV, AMF, and measured water vapour absorption (WVA) is captured by a function of the form f(x, y) = a x + b + c y$^2$ + d y + e x y, where x = WVA and y = AMF.
Figure \ref{fig:iwvfit}b illustrates the best fit of this model, based on 30-minute averages of all available clear-sky data during FESSTVaL for sensor 11 at Falkenberg compared to the microwave radiometer.
Measurement uncertainties are based on the standard deviation within the 30-minute windows.

Since there is a limitation to the accuracy of the spectra from sensor to sensor and within sensors (Section \ref{sec:frosterror}), we find no gain by training the model from data of all sensors together, or calibrating sensors individually as we do for GHI, likely to due over-fitting.
Instead, we apply the best fit based on sensor 11 at FESSTVaL, and use the uncertainty in the ratio between spectral bands as an error estimation.
Similar to broadband irradiance estimates, one could fine-tune the calibration on a case-specific basis if a high quality reference is available.
We estimate uncertainty by taking the standard deviation between all sensors in a network, which results in $\pm$ 0.5 kg m$^{-2}$ for sufficiently high solar elevation angles ($\alpha$ $>$ 30$^{\rm o}$), and increases towards sunset and sunrise.

\subsection{Sub-mesoscale water vapour variability at FESSTVaL}
\begin{figure}[ht]
    \centering
    \includegraphics[width=\textwidth]{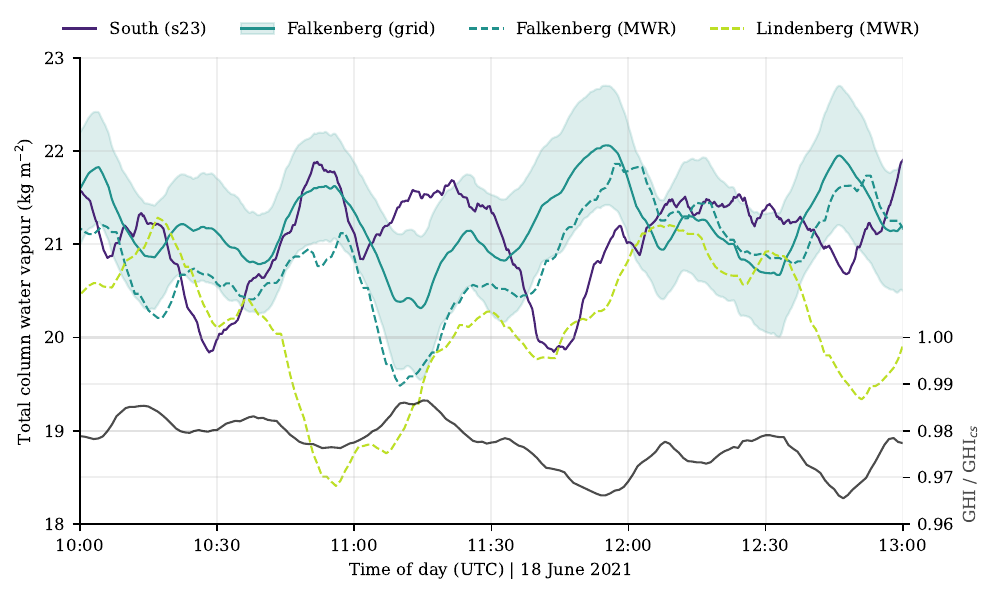}
    \caption{\textbf{Time series of total column water vapour (TCWV) from a south to north transact in the FESSTVaL campaign area.} FROST-derived TCWV is compared to high quality microwave radiometer measurements, both resampled to 10 minute moving averages. The shading is the FROST grid standard deviation. The south to north transact (south - Falkenberg - Lindenberg) is approximately 10 km, with Falkenberg in the middle (see also Figure \ref{fig:gridf}b), and the wind is $\approx$ 10 m s$^{-1}$ from the south. The black line is the Falkenberg global horizontal irradiance relative to clear-sky, also with a 10 minute moving average.}\label{fig:iwvfesstval}
\end{figure}

June 18 at FESSTVaL was a clear-sky, warm summer day (maximum 2 m temperature of 31 $^{\rm o}$C).
There are intra-hourly variations in TCWV of 1-3 kg m$^{-2}$, as illustrated in Figure \ref{fig:iwvfesstval} for various locations within the larger campaign area (Figure \ref{fig:gridf}b). 
With a predominantly southerly wind of $\approx$ 9 m s$^{-1}$ average over the lower troposphere, we tried to track moisture patterns across a 10 km south-north transect in the FESSTVaL domain (Figure \ref{fig:gridf}b), south of Falkenberg to Lindenberg, at a 10 and 20 minute time lag respectively.
The southern and Falkenberg measurements are based on the TCWV derived from our instruments, with at Falkenberg and Lindenberg two high quality measurements taken from the microwave radiometers.
While our instruments and the microwave radiometers at Falkenberg are in agreement on the local variations (despite a bias of $\sim$ 0.5 kg m$^{-2}$), there appears to be no correlation between the southern (s23) and northern (Lindenberg) location (Figure \ref{fig:iwvfesstval}).
We therefore think that the temporal variability in TCWV is more locally driven by turbulence structures in the convective boundary layer rather than the advection of (sub)-mesoscale horizontal patterns in moisture.
Observed variations in GHI at Falkenberg show a clear anti-correlation with TCWV, +1 mm TCWV $\simeq$ -1 \% GHI. 
This can not be explained by the direct effect of TCWV variations alone, which would be closer to -0.1 \% GHI for +1 mm of water vapour, and might point to buoyant plumes in the boundary layer that carry extra aerosols from the surface.
The more subtle differences in variability between irradiance and microwave radiometer based measurements may be explained by the different footprint of FROST sampling different moisture structures, given its horizontal component in the diagonal cross-section through the atmosphere (Figure \ref{fig:scheme_rad}a).

\subsection{Variability in water vapour at LIAISE}
\begin{figure}[ht]
    \centering
    \includegraphics[width=\textwidth]{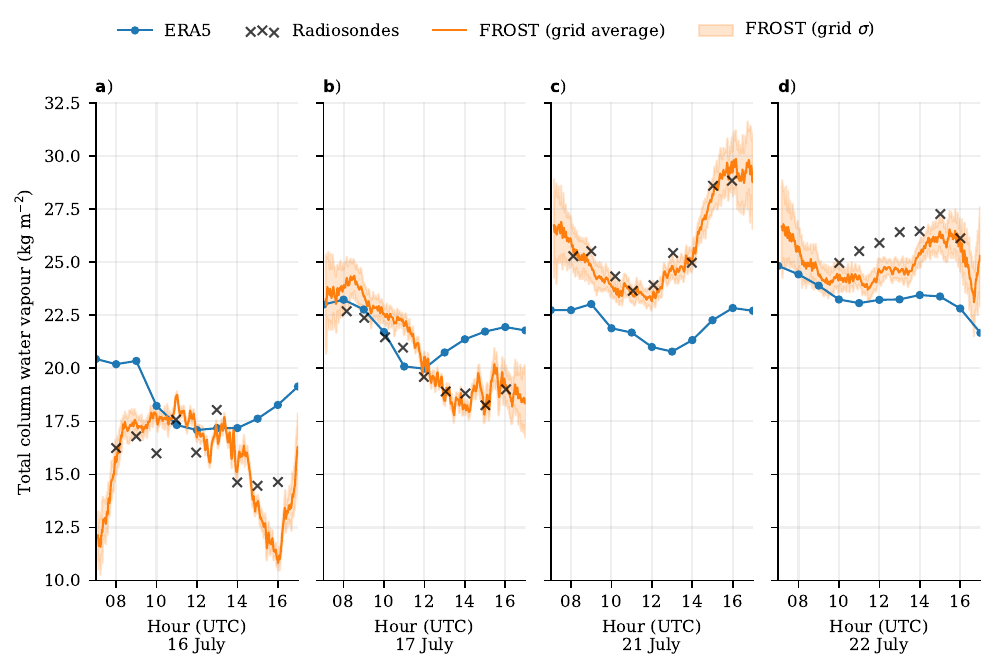}
    \caption{\textbf{Time series of total column water vapour (TCWV) at La Cendrosa during LIAISE.} Irradiance based measurements are compared to ERA5 (interpolated to La Cendrosa), and hourly soundings of La Cendrosa combined with Els Plans. The uncertainty range of the FROST measurements is the standard deviation among sensors in the network. The four dates are ideal clear-sky days and official Intensive Observation Periods.}
    \label{fig:iwvliaise}
\end{figure}

During the LIAISE campaign, most days were clear-sky (Figure \ref{fig:campaignwx}), which gives ample data to study intra-day variability in atmospheric moisture.
The campaign area is characterized by complex meteorology due to a combination of topography, strong heterogeneity in surface fluxes at various scales due to irrigation (Bowen ratio between 0.01 and 30, \cite{mangan_surface-boundary_2023}), and late afternoon south-eastern sea breeze dynamics with varying timing and strength.
This collection of complex factors affecting local weather is not captured by high resolution weather models, and even less so by the one ERA5 grid cell covering the campaign area.
Measuring TCWV helps to identify internal boundary layers and moisture plumes at the local scale to synoptic scale advection, that both contribute to variability.
We apply the best fit derived from FESSTVaL (Figure \ref{fig:iwvfit}) to the sensor network at the irrigated La Cendrosa site.
Figure \ref{fig:iwvliaise} shows the resulting time series for four (clear-sky) IOP days compared against ERA5 and hourly radiosondes, using a 3 minute moving average to get a clear signal.
FROST-derived TCWV time series and soundings do not at all agree with the magnitude and trends of ERA5, which highlights the difficulty (coarse) models have with the complex meteorology in the LIAISE domain.
Radiosondes and our spectrally derived TCWV time series are in much better agreement overall, and provides a good validation of our calibration methodology.
Derivation of TCWV from radiosondes is precarious, though, since the hourly boundary layer radiosondes at La Cendrosa \citep{price_radiosondes_2023} only reach to about 1.5 - 4 km and need to be supplemented with hourly tropospheric radiosondes from Els Plans \citep{canut_radiosondes_2022}, a non-irrigated location 14.1 km to the south-east.
The atmospheric conditions in Els Plans and La Cendrosa converge above their respective local boundary layers to a regional atmospheric profile, described by a blending height of approximately 1.5 km \citep{mangan_surface-boundary_2023}.
For each sounding in La Cendrosa, we supplement its information with the mid to upper tropospheric data gathered from the closest (in time) Els Plans sounding, as schematically illustrated in Figure \ref{fig:scheme_iwv}b).
Timing and footprint differences between instantaneous spectrum-derived TCWV and hourly combined soundings are, we believe, the main reason for differences between their measurements.

\begin{figure}[ht]
    \centering
    \includegraphics[width=\textwidth]{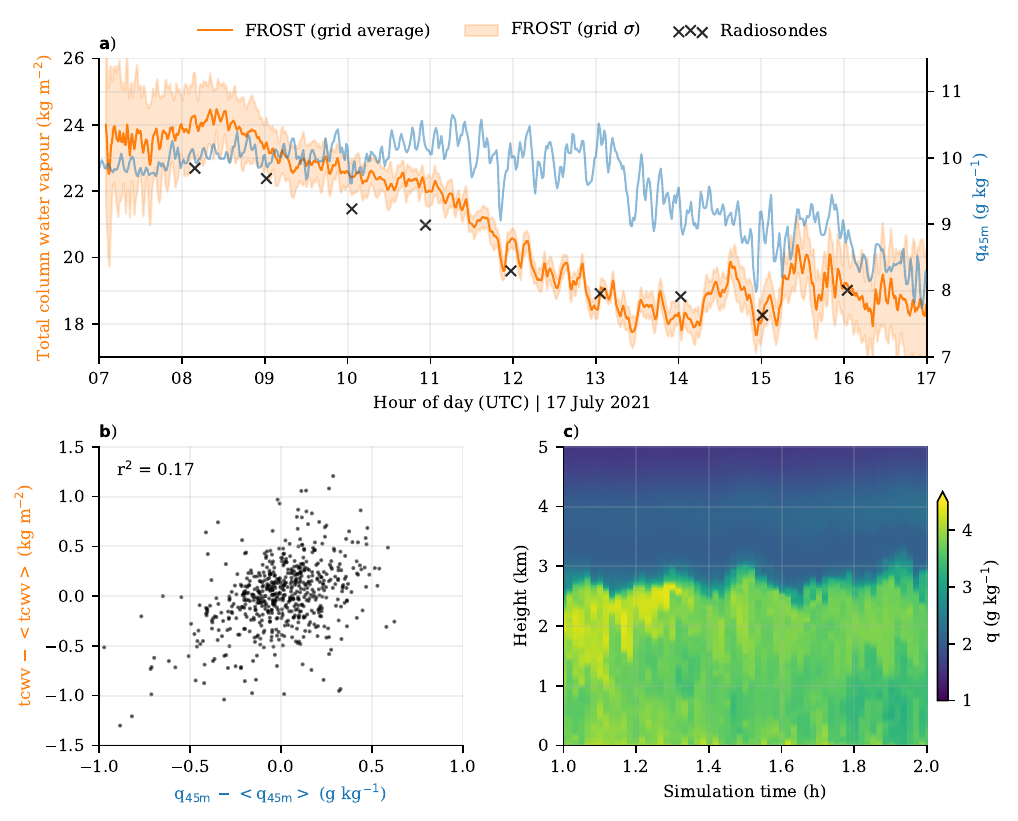}
    \caption{\textbf{Moisture variability of total column compared to 45 meters at La Cendrosa.} \textbf{(a)} shows a detailed time series of the same data as Figure \ref{fig:iwvliaise}b, combined with specific humidity at 45 meters above ground level. The correlation between deviations from hourly mean values for both total column water vapour and specific humidity at 45 meters are shown in \textbf{(b)}. To illustrate the difference between total column and single point moisture, \textbf{(c)} shows a (time, height) cross section of a convective boundary layer from large eddy simulation.}
    \label{fig:iwv_vs_q}
\end{figure}

In terms of variability, there seem to be various time scales at play. 
On hourly time scales, trends in TCWV can be up to 5 kg m$^{-2}$ h$^{-1}$, whereas at the minute time scales the constant fluctuations do not exceed 1 kg m$^{-2}$.
The latter we believe to be driven by moist boundary layer thermals and subsequent dry air entrainment.
Daily cycles of boundary layer drying through local advection \citep{mangan_surface-boundary_2023} and synoptic scale advection of air masses (captured by ERA5), explain the multi-hour and day to day variations, respectively.
Figure \ref{fig:iwv_vs_q} illustrates a detailed look at short term variability of TCWV, including an analysis in which we attempt to correlate variability of measured specific humidity at 45 meters ($\rm q_{45\rm m}$) in the well mixed boundary layer to variation in TCWV. 
Both time series are re-sampled to a common resolution of 2 minutes.
To exclude hourly or daily trends, we define variations in TCWV or $\rm q_{45\rm m}$ as the deviations with respect to their 60 minute moving average.
The boundary layer contains a significant portion of the total vertical moisture, but the resulting correlation is only very weak, with an r$^2$ of 0.17 (Figure \ref{fig:iwv_vs_q}b), and thus can't be explained by local moist or dry plumes.
Rather, the TCWV from FROST has larger scale variations due to its diagonal cross-sectional footprint, as compared to a single point measurement.
Figure \ref{fig:iwv_vs_q}c shows the vertical structure of $\rm q$ for a single point in an idealised convective boundary layer simulation in north-eastern Spain, where the complex turbulence structures illustrate why variations between $\rm q_{45\rm m}$ and TCWV do not correlate.
FROST is thus able to capture moisture variability that is representative for the kilometre scale and above rather than at individual field level. 
In an improved version of FROST, we have an improved signal to noise performance and thus require shorter time averaging, which may improve its ability to capture more small scale variations.
This could offer a flexible, low-cost alternative to a microwave radiometer or soundings.

\clearpage

\section{Conclusions and Outlook}\label{sec:conclusions}
Using low-cost radiometers calibrated against high-quality reference stations, we have gathered two high-resolution, spectrally resolving datasets of surface solar irradiance at cloud-scale. 
In combination with supplementary observations and simulations, such as cloud cameras at FESSTVaL, soundings at LIAISE, and clear-sky spectra, we are able to derive insights into spatial patterns of surface irradiance caused by various types of clouds and by atmospheric moisture variability.
This work demonstrates how low-cost instruments can provide accurate and detailed spatial measurements and, given their flexible setup, can be an effective addition to field campaigns, particularly in areas for which deployment of expensive, heavy equipment is difficult.

We analysed a case of boundary layer cumulus, mid level altocumulus, and high level cirrus.
All three presented cases have distinctly different surface irradiance patterns, spatial scales scales of variability (50 m to 30 km), cloud types, direct/diffuse partitioning, spectral irradiance, and mechanisms through which these patterns are formed. 
The contribution of various scattering mechanisms is difficult to quantify, but by comparing changes between visible blue and red wavelengths, we think the enhanced irradiance next to the shadow of a cumulus has a large contribution of locally forward-scattered light through the cumulus cloud edge region, whereas further away from a shadow the enhanced radiation comes mostly from cloud sides of the overall cloud field.
For the mid and high level cloud cases, conditions of reduced irradiance are more blue and enhanced more red compared to clear-sky, though the cause of these changes in spectral irradiance is inconclusive.

None of these commonly occurring cases are well-represented by state-of-the-art numerical weather prediction, nor by most cloud-resolving models in academic setups, due to simplifications in radiative transfer calculations.
In particular altocumulus is potent in creating strong, localised peaks, and may be underrepresented in the field of 3D radiative transfer research.
We have only focused on cloud cases that occur in isolation, to make interpretation easier, but often clouds of varying type occur simultaneously and the effects may not simply be additive. 
Despite some limitations of our sensor, we are also able to capture variability in the irradiance spectrum that arises from significant changes in atmospheric water vapour in clear-sky conditions.
These local variations are often larger than synoptic scale moisture advection, but correlate with variations in GHI in the order of a percent, and illustrate heterogeneity in moisture fluxes and optical properties of a cloud-free boundary layer.

The presented datasets provide observations of cloud- and moisture-driven irradiance variability that can guide the development of radiative transfer variability parameterisations and better constrain the input for land-surface, photosynthesis, or dynamic vegetation models that are currently driven by incorrect irradiance distributions or spectral properties.
A more comprehensive understanding of cloud-driven irradiance variability will require analyses of many more clouds and irradiance patterns in datasets such as these, possibly aided by cloud resolving models and accurate 3D radiative transfer solvers to quantitatively characterise the mechanisms.

\section{Open Data}\label{sec:opendata}
All data measured with FROST sensors at FESSTVaL and LIAISE are published open-access datasets.
These include ready to use calibrated quality controlled data and the raw instrument data, which requires pyranometer calibration references:
\begin{itemize}
    \item Radiometer data FESSTVaL \citep{mol_fesstval_2023}: \url{https://doi.org/10.25592/uhhfdm.10272}
    \item Radiometer data LIAISE \citep{mol_liaise_2023}: \url{https://doi.org/10.5281/zenodo.7966437}
\end{itemize}

\noindent The code for calibration of raw data and analyses presented in this work, the video of Section \ref{sec:video}, and the libRadtran input files, are available at \url{https://zenodo.org/records/10159129}.
La Cendrosa solar irradiance is not yet available at the time of writing, but will be available on the LIAISE database: \url{https://liaise.aeris-data.fr/page-catalogue/?uuid=d9608a55-b836-427b-a186-e007462012b9}.

\section{Author contributions}
W.M. has designed and operated the sensor network measurements at both campaigns, processed and analysed the data, and wrote the manuscript. 
B.H. has designed and calibrated the sensors, and helped with the measurements at FESSTVaL. 
M.V. has helped with sensor maintenance at FESSTVaL, MR. M. and O.H. have led the Dutch team at LIAISE and helped with analyses of LIAISE observations.
C.v.H. has helped with designing the fieldwork, sensor deployment at FESSTVaL, analysing results, and writing the final manuscript.
All authors have contributed to the final manuscript.

\section{Acknowledgements}
Our thanks goes out to the FESSTVaL and LIAISE campaign organisers and host institutes for enabling us to gather observations alongside many others, without which this work would not have been possible. 
W.M., B.H., M.V, and C.v.H, acknowledge funding from the Dutch Research Council (NWO) (grant: VI.Vidi.192.068).

\section{Supporting Information}
This manuscript contains additional figures and a table available as supporting information.

\bibliography{zotero_library}

\clearpage

\end{document}